\documentstyle[epsfig]{mn}

\newif\ifAMStwofonts

\ifoldfss
  \ifCUPmtlplainloaded \else
    \NewTextAlphabet{textbfit} {cmbxti10} {}
    \NewTextAlphabet{textbfss} {cmssbx10} {}
    \NewMathAlphabet{mathbfit} {cmbxti10} {} 
    \NewMathAlphabet{mathbfss} {cmssbx10} {} 
  \fi
  \ifAMStwofonts
    \ifCUPmtlplainloaded \else
      \NewSymbolFont{upmath} {eurm10}
      \NewSymbolFont{AMSa} {msam10}
      \NewMathSymbol{\upi}     {0}{upmath}{19}
      \NewMathSymbol{\umu}     {0}{upmath}{16}
      \NewMathSymbol{\upartial}{0}{upmath}{40}
      \NewMathSymbol{\leqslant}{3}{AMSa}{36}
      \NewMathSymbol{\geqslant}{3}{AMSa}{3E}

    \fi
  \fi
\fi 

\ifnfssone
  \newmathalphabet{\mathit}
  \addtoversion{normal}{\mathit}{cmr}{m}{it}
  \addtoversion{bold}{\mathit}{cmr}{bx}{it}
  \newmathalphabet{\mathbfit} 
  \addtoversion{normal}{\mathbfit}{cmr}{bx}{it}
  \addtoversion{bold}{\mathbfit}{cmr}{bx}{it}
  \newmathalphabet{\mathbfss} 
  \addtoversion{normal}{\mathbfss}{cmss}{bx}{n}
  \addtoversion{bold}{\mathbfss}{cmss}{bx}{n}
  \ifAMStwofonts
    \ifCUPmtlplainloaded \else
      \UseAMStwoboldmath
      \makeatletter
      \new@mathgroup\upmath@group
      \define@mathgroup\mv@normal\upmath@group{eur}{m}{n}
      \define@mathgroup\mv@bold\upmath@group{eur}{b}{n}
      \edef\UPM{\hexnumber\upmath@group}
      \new@mathgroup\amsa@group
      \define@mathgroup\mv@normal\amsa@group{msa}{m}{n}
      \define@mathgroup\mv@bold\amsa@group{msa}{m}{n}
      \edef\AMSa{\hexnumber\amsa@group}
      \makeatother
      \mathchardef\upi="0\UPM19
      \mathchardef\umu="0\UPM16
      \mathchardef\upartial="0\UPM40
      \mathchardef\leqslant="3\AMSa36
      \mathchardef\geqslant="3\AMSa3E
    \fi
  \fi
\fi 

\ifnfsstwo
  \DeclareMathAlphabet{\mathbfit}{OT1}{cmr}{bx}{it}
  \SetMathAlphabet\mathbfit{bold}{OT1}{cmr}{bx}{it}
  \DeclareMathAlphabet{\mathbfss}{OT1}{cmss}{bx}{n}
  \SetMathAlphabet\mathbfss{bold}{OT1}{cmss}{bx}{n}
  \ifAMStwofonts
    \ifCUPmtlplainloaded \else
      \DeclareSymbolFont{UPM}{U}{eur}{m}{n}
      \SetSymbolFont{UPM}{bold}{U}{eur}{b}{n}
      \DeclareSymbolFont{AMSa}{U}{msa}{m}{n}
      \DeclareMathSymbol{\upi}{0}{UPM}{"19}
      \DeclareMathSymbol{\umu}{0}{UPM}{"16}
      \DeclareMathSymbol{\upartial}{0}{UPM}{"40}
      \DeclareMathSymbol{\leqslant}{3}{AMSa}{"36}
      \DeclareMathSymbol{\geqslant}{3}{AMSa}{"3E}
    \fi
  \fi
\fi 

\ifCUPmtlplainloaded \else
  \ifAMStwofonts \else 
    \def\upi{\pi}
    \def\umu{\mu}
    \def\upartial{\partial}
  \fi
\fi

\title
[Observability of secondary Doppler peaks]
{Observability of secondary Doppler peaks in the CMBR power spectrum
by experiments with small fields}
\author[M.P. Hobson, Joao Magueijo]{M.P. Hobson$^1$, Joao Magueijo$^{1,2}$ \\
$^1$Mullard Radio Astronomy Observatory,
Cavendish Laboratory, Madingley Road, Cambridge, CB3 0HE, U.K.\\
$^2$ Department of Applied Mathematics and Theoretical
Physics, University of Cambridge, Cambridge CB3 9EW, U.K.}
\date{Accepted ???. Received ???; in original form ???}
\pagerange{\pageref{firstpage}--\pageref{lastpage}}
\pubyear{1996}

\begin{document}
\maketitle
\label{firstpage}

\begin{abstract}
We investigate the effects of finite sky coverage on the spectral resolution
$\Delta\ell$ in the estimation of the CMBR angular power spectrum $C^{\ell}$.
A method is developed for obtaining quasi-independent estimates of the 
power spectrum, and the cosmic/sample variance of these
estimates is calculated. The effect of instrumental noise is also considered
for prototype interferometer and single-dish experiments.
By proposing a statistic for the detection of 
secondary (Doppler) peaks in the CMBR power spectrum, we then compute 
the significance level at which such peaks may be detected for a large
range of model CMBR experiments. 
In particular, we investigate experimental design features required
to distinguish between competing cosmological theories,
such as cosmic strings and inflation,  by establishing whether
or not secondary peaks are present in the CMBR power spectrum.
\end{abstract}


\section{Introduction}
The cosmic microwave background radiation (CMBR) is one of the most promising
links between astronomical observation and cosmological theory. A significant
amount of experimental data already exists and further observational
progress is expected in the near future. The experimental 
success in this field  has prompted a large
amount of theoretical effort on two closely
related fronts. Firstly, theorists strive to assess the impact
current  observations have on theories of the early Universe. 
Secondly, the design of future experiments is guided
by what are believed to be crucial tests of our theoretical prejudices.

A particularly active field of research in CMBR physics are the so-called
Doppler peaks (Hu \& Sugiyama 1995a,b). 
These consist of a series of oscillations in the 
angular power spectrum of CMBR fluctuations $C^{\ell}$ predicted for most
inflationary models. They are predicted in the multipole
range $100\la \ell \la 1500$, corresponding 
to angular scales $0.05 \la \theta\la 1$ degrees.
Experimental measurement of the Doppler peaks'
positions and heights would fix at least some combinations of  
cosmological parameters (e.g. $H_0$,
$\Omega_0$ etc.) which are left free in inflationary models (Jungmann
1995). Future experimental projects
are usually designed with these goals in mind.
In particular, models with $\Omega_0 =1$ are of special interest.

In this paper we shall consider another theoretical context.
For a long time inflation  (Steinhardt 1995), 
and topological defects (Kibble 1976; Vilenkin \& Shellard 1994),
have stood as conceptually opposing alternative scenarios for 
structure formation in the early Universe. 
It was shown by Albrecht et al (1996) and
Magueijo et al (1996),
that qualitative aspects of Doppler peaks should reflect this
conceptual opposition. In particular the absence of secondary Doppler
peaks was proved to be a robust prediction for a large section of defect
theories. Interestingly enough this statement relies only
on the role played by causality
and randomness in these theories. Cosmic strings were 
unambiguosly shown to fall in this category. 
However, this may or may not be the case for textures  
(Crittenden \& Turok 1995; and Durrer et al 1995). 
Therefore it appears that determining 
whether or not there are secondary Doppler
peaks in the CMBR power spectrum
offers an important alternative motivation 
for experimental design and data analysis. Answering this
question is a far less demanding experimental challenge,
which nevertheless will have a dramatic impact on our 
understanding of the Universe (Magueijo \& Hobson 1996, 
Albrecht \& Wandelt 1996).

We address this issue by proposing in Section~\ref{peakstat}
a statistic for detecting 
secondary oscillations, and studying how it fares for signals
coming from various models, when measured by using different experimental 
strategies. The result is encoded in a detection function $\Sigma$ 
telling us to within how many sigmas we can claim a detection of secondary
oscillations, given a particular model and experiment. 
We consider signals coming from standard CDM (sCDM),
and an open CDM model (henceforth called stCDM) which is 
tuned to confuse inflation and cosmic strings in all but
the issue of secondary oscillations. 
In Secs.~\ref{sfobs}, \ref{cv}, and \ref{noise} we set up a framework
for computing errors in power spectra estimates. We consider
errors  resulting from  
spectral resolution limitations due to finite sky coverage,
cosmic/sample variance, instrumental noise and foreground subtraction.
We consider a large parameter space of experiments including 
single-dish experiments
and interferometers. For single-dish experiments we allow 
the beam size, sky coverage, and detector noise to vary. For
interferometers we take as free parameters
the primary beam, number of fields,
and detector noise.  We then use this framework to compute the 
detection function $\Sigma$ for sCDM and stCDM in this large
class of experiments. 

The results obtained are given in Sections~\ref{peakstat} 
and \ref{cs}, and provide experimental guidance in two different ways.
Firstly they allow the choice of an ideal scanning strategy (choice
of resolution and sky coverage) given
a constraint such as finite funding (albeit 
disguised in a more mathematical form). 
Secondly, one may compute the expected value of the detection, 
assuming ideal scanning, as a function of available funding.
This provides lower bounds on experimental conditions
for a meaningful detection
as well as an estimate of how fast detections will improve thereafter.
We summarize
the main results in the Section~\ref{conc}.

\section{Observations of small fields}
\label{sfobs}

CMBR temperature fluctuations are usually described as an expansion in
spherical harmonics
\[
\frac{\Delta T(\bmath{\hat{x}})}{T}   
= \sum_{\ell=0}^{\infty} \sum_{m=-\ell}^{\ell}
a_{\ell m} Y_{\ell m} (\bmath{\hat{x}}),
\]
from which we define the CMBR power spectrum $C^{\ell} = \langle |a_{\ell
m}|^2\rangle$. Note that in this paper we write ensemble average quantities
with the indices upstairs, and denote {\em observed} values of such variables 
in any given experiment with the indices downstairs.

If one is looking only at small patches of the sky, the spherical harmonic
analysis is awkward to apply, and it is more convenient to use Fourier
analysis. For that purpose we perform a stereographic projection of the
celestial sphere onto a tangent plane at the centre of our small patch, so
that circles with colatitude $\theta$ are mapped onto circles of radius
$r=2\tan (\theta/2)$ on the plane.  We can then describe the CMBR fluctations
$\Delta T(\bmath{x})/T$ on this plane by its Fourier transform. We use the
conventions
\[
{\Delta T(\bmath{x})\over T}
={\int {d^2\bmath{k}\over 2\pi}a(\bmath{k})e^{i\bmath{k}\cdot\bmath{x}}}\; ,
\]
and
\[
a(\bmath{k})
={\int {d^2\bmath{x}\over 2\pi}{\Delta T(\bmath{x})\over T}
e^{-i\bmath{k}\cdot\bmath{x}}}.
\]

If we denote the Fourier transform of the autocorrelation function of the
$\Delta T(\bmath{x})/T$ field by $C(\bmath{k})/(2\pi)$ then the covariance of
the Fourier modes $a(\bmath{k})$ is given by
\begin{equation}\label{pkraw}
\langle a(\bmath{k})a^*(\bmath{k}') \rangle
=C(k)\delta(\bmath{k}-\bmath{k}'),
\end{equation}
where the angle brackets denote ensemble averages, * denotes complex
conjugation and $k=|\bmath{k}|$. We note that, due to the requirement of
rotational invariance, $C(\bmath{k})=C(k)$. From (\ref{pkraw}),
the raw modes $a(\bmath{k})$
are therefore independent Gaussian random variables with variance $C(k)$. 
In calculations concerning small patches of the sky $C(k)$ can be
obtained by interpolating the $C^{\ell}$ coefficients in the spherical
harmonic expansion with $|\bmath{k}|=\ell$ (Bond \& Efstathiou 1987).  An
improvement can be obtained with the prescription $C(k)=C[\ell(k)]dk/dl$ with
$k(\ell)=\pi\cot(\pi/\ell)$.

We wish to study the statistical properties of the $a(\bmath{k})$ modes as
seen by observations of a small field with a given observing beam.  Let us
describe the field by a window $W(\bmath{x})$ and the observing beam by
$B(\bmath{x})$. The sampled temperature map for a single-dish experiment is
then
\[
{\Delta T_s(\bmath{x})\over T}
=\left[{\Delta T(\bmath{x})\over T}\star B(\bmath{x})\right]W(\bmath{x}),
\]
where $\star$ denotes convolution. The {\em sampled} Fourier modes 
$a_s(\bmath{k})$ are therefore given by
\begin{eqnarray}
a_s(\bmath{k}) 
& = &
[a(\bmath{k})\widetilde{B}(\bmath{k})]\star\widetilde{W}(\bmath{k}) 
\nonumber \\
& = & \int {d^2\bmath{k}'\over 2\pi} \widetilde{W}(\bmath{k}-\bmath{k}')
a(\bmath{k}')\widetilde{B}(\bmath{k}'), 
\label{as}
\end{eqnarray}
where $\widetilde{W}(\bmath{k})$ and $\widetilde{B}(\bmath{k})$ denote the
Fourier transforms of the window and observing beam respectively. 

For interferometers, however, we must make a slight modification since in this
case the Fourier domain is sampled directly.  An interferometer samples the
Fourier transform of the product of the sky temperature fluctuations and the
primary beam of the antennas $W(\bmath{x})$ (which corresponds to a window).
The positions of these samples (or visibilities)
in the Fourier domain (or $uv$-plane) are
determined by the physical positions of the antennas and the direction to the
patch of sky being observed, and lie on a series of curves (or
$uv$-tracks). If we denote these curves by the function
$\widetilde{B}(\bmath{k})$, which equals unity where the Fourier
domain in sampled
and equals zero elsewhere, then the sampled Fourier modes for an
interferometer are given by
\begin{eqnarray}
a_s(\bmath{k}) 
& = & [a(\bmath{k})\star
\widetilde{W}(\bmath{k})]\widetilde{B}(\bmath{k}) 
\nonumber \\
& = & \widetilde{B}(\bmath{k}) 
\int {d^2\bmath{k}'\over 2\pi} 
\widetilde{W}(\bmath{k}-\bmath{k}') a(\bmath{k}'). 
\label{as2}
\end{eqnarray}
The inverse Fourier transform of $\widetilde{B}(\bmath{k})$ is the synthesised
beam of the interferometer $B(\bmath{x})$. The $a_s(\bmath{k})$ modes
are often inverse Fourier transformed to make a sky map. This sampled sky 
temperature distribution for an interferometer is
\[
{\Delta T_s(\bmath{x})\over T}
=\left[{\Delta T(\bmath{x})\over T}W(\bmath{x})\right]\star B(\bmath{x}),
\]
which is the convolution of the synthesised beam with the product of the
sky and the primary beam (or window).

From (\ref{pkraw}), (\ref{as}) and (\ref{as2}) one can then derive the 
covariance matrix of the sampled modes for each experiment 
(Hobson, Lasenby \& Jones 1995). For a single-dish observation we find
\[
\langle a_s(\bmath{k})a_s^*(\bmath{k}')\rangle =
\]
\begin{equation}
~~~~\int {d^2\bmath{k}''\over (2\pi)^2} 
\widetilde{W}(\bmath{k}-\bmath{k}'')
\widetilde{W}^*(\bmath{k}'-\bmath{k}'')
C(\bmath{k}'')|\widetilde{B}(\bmath{k}'')|^2,
\label{covmat}
\end{equation}
whereas for an interferometer
\[
\langle a_s(\bmath{k})a_s^*(\bmath{k}')\rangle =
\]
\begin{equation}
~~~~|\widetilde{B}(\bmath{k})|^2
\int {d^2\bmath{k}''\over (2\pi)^2} 
\widetilde{W}(\bmath{k}-\bmath{k}'')
\widetilde{W}^*(\bmath{k}'-\bmath{k}'')
C(\bmath{k}'').
\label{covmat2}
\end{equation}
Ignoring, for the moment, the effects of the observing beam $B(\bmath{x})$ in
each case, we see that finite sky coverage [described by the window
$W(\bmath{x})$] renders the observed $a_s(\bmath{k})$ {\em dependent} random
variables. Roughly speaking, if $L$ is the typical size of the field then a
correlation length $\xi \approx 1/L$ is introduced in the $\bmath{k}$-plane.

\subsection{The sampled power spectrum}
\label{tsps}

The two-dimensional sampled power spectrum may be defined in terms of the
marginal variances of the sampled modes as 
\[
{\sf C}^s(\bmath{k})=\langle a_s(\bmath{k})a_s^*(\bmath{k})\rangle
\]
From (\ref{covmat}) and (\ref{covmat2}) we see that ${\sf C}^s(\bmath{k})$ is
given for single-dish experiments by
\begin{eqnarray}
{\sf C}^s(\bmath{k}) & = & \int \frac{d^2\bmath{k}'}{(2\pi)^2}
|\widetilde{W}(\bmath{k}-\bmath{k}')|^2 C(\bmath{k}')
|\widetilde{B}(\bmath{k}')|^2 \nonumber \\
& = & \left[C(\bmath{k})|\widetilde{B}(\bmath{k})|^2\right]\star
|\widetilde{W}(\bmath{k})|^2,
\label{css}
\end{eqnarray}
and for interferometers by
\begin{eqnarray}
{\sf C}^s(\bmath{k}) & = & 
|\widetilde{B}(\bmath{k})|^2 \int \frac{d^2\bmath{k}'}{(2\pi)^2}
|\widetilde{W}(\bmath{k}-\bmath{k}')|^2 C(\bmath{k}') \nonumber \\
& = & |\widetilde{B}(\bmath{k})|^2\left[C(\bmath{k})\star
|\widetilde{W}(\bmath{k})|^2\right].
\label{csi}
\end{eqnarray}
Ignoring, for the moment, the effects of the observing beam $B(\bmath{x})$
[by setting its Fourier transform $\widetilde{B}(\bmath{k})$ to unity in
(\ref{css}) and (\ref{csi})], we see that in both cases the two-dimensional
sampled power spectrum is the convolution of the underlying
CMBR power spectrum $C(k)$ with $|\widetilde{W}(\bmath{k})|^2$. We also note
that in general ${\sf C}^s(\bmath{k})$ is {\em not} circularly symmetric. We
therefore define the one-dimensional sampled power spectrum as the
azimuthal average $C^s(k)=\langle{\sf C}^s(k,\theta)\rangle_{\theta}$, where
$\theta$ is the azimuthal angle in the Fourier domain.

In order to provide examples of the effects of the window on the sampled 
power spectrum, we now consider two simple windows: a square window and
a Gaussian window. A square window is a toy model for single-dish experiments.
If the square window has sides of length $L$, then its Fourier transform is
given by
\begin{equation}\label{wink}
\widetilde{W}(\bmath{k})={L^2\over 2\pi}j_0\left({k_x L\over 2}\right)
j_0\left({k_y L\over 2}\right),
\end{equation}
where $j_0(x)=\sin x/x$ is the spherical Bessel function of order zero.
A Gaussian window normalized to one at its peak is a toy model for an
inteferometer primary beam, and is given by
\[
W(\bmath{x})=\exp\left(-\frac{x^2}{2\sigma_w^2}\right),
\]
which has the Fourier transform
\[
\widetilde{W}(\bmath{k})=\sigma_w^2\exp\left(-\frac{k^2\sigma_w^2}{2}\right).
\]
We note that both the square and Gaussian windows are real, even functions, 
and hence so are there Fourier transforms. 

Using these windows we may now calculate the
sampled power spectrum for various experiments (assuming for the moment that
$\widetilde{B}(\bmath{k})=1$), given an underlying CMBR power
spectrum. We choose
the underlying spectrum to be that predicted by the standard inflation/CDM 
scenario with $\Omega_0=1$, $h_0=0.5$, and $\Omega_b=0.05$ (which we shall
call sCDM). In order to accentuate the effects of the windows we first
consider small fields.

\begin{figure}
\centerline{\epsfig{
file=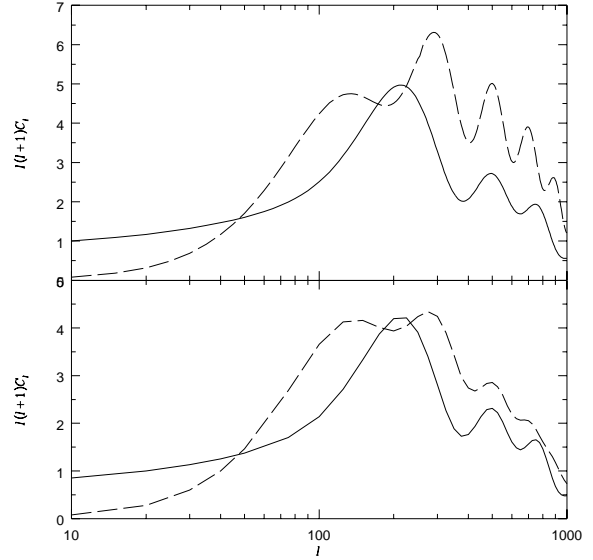,
width=8cm}}
\caption{(Top) The sampled power spectrum ${\sf C}^s(k,0)$ (dashed line) 
for a single beam experiment with a square window of size $L=1.8$ degrees and
resolution $k_b=1000$, 
compared to the underlying sCDM power spectrum (solid line) as observed by
an all sky experiment with the same resolution. (Bottom) The same
as above but for the azimuthal average sampled power spectrum
$C^s(k)=\langle {\sf C}^s(k,\theta)\rangle_{\theta}$.}
\label{fig1}
\end{figure}

As a model single-dish experiment we consider a square field of size
$L=1.8$ degrees observed with a Gaussian beam with $\sigma_w=0.06$ degrees
(or FWHM equal to 0.14 degrees), so that
\begin{equation}
\widetilde{B}(k)=\exp\left(-\frac{k^2}{2k_b^2}\right),
\label{ftbeam}
\end{equation}
where $k_b=1000$. 
For a square window the two-dimensionsal sampled power spectrum is not
circularly symmetric, and so ${\sf C}^s(k,\theta)$ is a function of $\theta$.
If we fix $\theta=\theta_0$ for any value of $\theta_0$ in the range
$[0,2\pi]$, we find that the spectrum starts as white-noise, i.e.
${\sf C}^s(k,\theta_0)=\lambda$ for some constant $\lambda$, up to
$k_{\rm min}\approx 2\pi/L$. From that point on, a spurious set of oscillations
with period $\approx 4\pi/L$ appear superposed on the raw spectrum.
This illustrated in Fig.~\ref{fig1} in which we plot ${\sf C}^s(k,0)$.
The spurious oscillations merely reflect the `ringing' of the Fourier
transform due to the sharp edges of the window. In the 1-dimensional
azimuthal average power spectrum $C^s(k)={\langle
{\sf C}^s(k,\theta)\rangle}_{\theta}$ 
these spurious oscillations are mainly
smoothed out, but so too are features in the raw spectrum on scales less than
$\Delta\ell\approx 4\pi/L$ (see Fig.~1\ref{fig1}).

It is well-known (e.g. Press et al 1994) that the spurious oscillations in
Fourier transforms can be suppressed by multiplying the 
data (sky map) by a window that goes to zero 
smoothly at its edges. In the context of CMB measurements, Tegmark (1996)
calculates the optimal window to use in various circumstances to obtain the
highest possible $\Delta \ell$ resolution in the sampled power spectrum.  For
a square field this window is a cosine `bell'
\begin{equation}
W(\bmath{x})=\cos\left(\frac{\pi x}{L}\right)\cos\left(\frac{\pi y}{L}\right),
\label{bellx}
\end{equation}
whose Fourier transform is
\begin{equation}\label{bell}
W(\bmath{k})={2\pi\over L^2}{\cos{(k_xL/2)}\over k_x^2-(\pi/L)^2}
{\cos{(k_yL/2)}\over k_y^2-(\pi/L)^2}.
\end{equation}
At any given point $k_0$ in the sampled power spectrum
this window minimises the variance (second central moment) 
$\langle (k-k_0)^2 \rangle$ of the convolving function 
$|\widetilde{W}(\bmath{k})|^2$. Interestingly, the more commonly encountered
Hann window
\[
W(\bmath{x})=\frac{1}{2}\left[1+\cos\left(\frac{2\pi x}{L}\right)\right]
\left[1+\cos\left(\frac{2\pi y}{L}\right)\right],
\]
minimises the fourth central moment $\langle (k-k_0)^4 \rangle$ of the
convolving function.

\begin{figure}
\centerline{\epsfig{
file=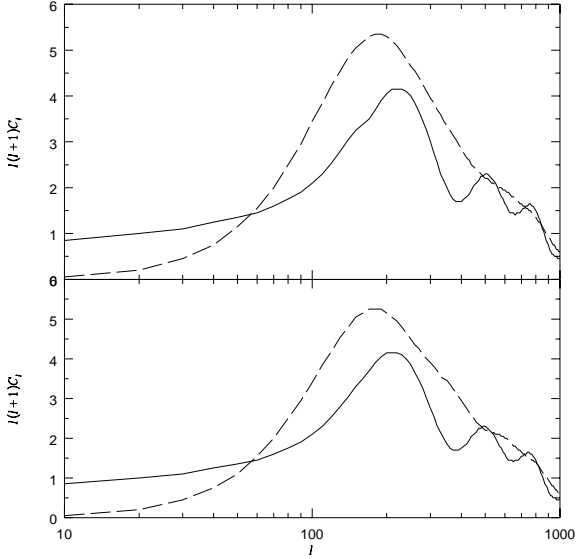,
width=8cm}}
\caption{As in Fig.~\ref{fig1} but for a square multiplied by a cosine bell.}
\label{fig2}
\end{figure}

Using the cosine bell window (\ref{bellx}), the spurious oscillations 
disappear from ${\sf C}^s(k,0)$ and $C^s(k)$, but for such a small field the 
features of the raw spectrum are still heavily smeared (see Fig.~\ref{fig2}).

\begin{figure}
\centerline{\epsfig{
file=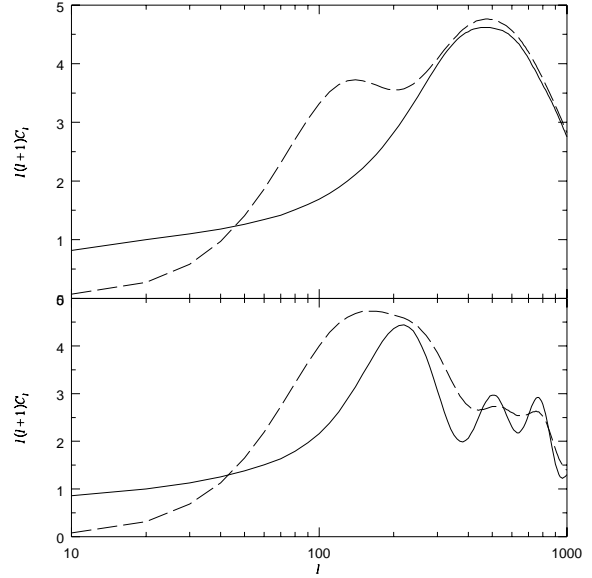,
width=8cm}}
\caption{(Top) The sampled power spectrum (dashed line) 
observed by an experiment with a Gaussian window with $\sigma_w=0.6$ degrees,
compared to the underlying CMBR spectrum predicted by cosmic strings (solid
line). (Bottom) The same as above, but for the sCDM spectrum.}
\label{fig3}
\end{figure}

For interferometers we may model the window (primary beam) as Gaussian.  In
Fig.~\ref{fig3} we show the effect on the sampled power spectrum
of a Gaussian window with $\sigma_w = 0.6$
degrees, which corresponds to a FWHM of 1.3 degrees.  Although for a real
interferometer observation, $\widetilde{B}(\bmath{k})=1$ only where the Fourier
domain is sampled (and is zero elsewhere), in order to illustrate the effect of
the window we set it equal to unity for all $\bmath{k}$.  
In Fig.~\ref{fig3}, we
again observe a white-noise tail up to $k_{\rm min} \approx 1/\sigma_w$, but
from then on the Gaussian window does not induce spurious oscillations on the
spectrum. The Gaussian window 
(which occurs naturally for inteferometer
observations) is in fact quasi-optimal, and the procedure suggested by Tegmark
(1996) makes a negligible difference to the spectral resolution of the sampled
power spectrum. However, this window is still too narrow since features in the
underlying spectrum on the scale of $\mbox{few}\times 1/\sigma_w$ 
are smoothed out.
Hence although Gaussian windows do not distort smooth spectra, 
they still erase
oscillations in oscillatory spectra if the window is too small. In
Fig.~\ref{fig3} we show that effect of the Gaussian window on CMBR
power spectra predicted by sCDM and cosmic strings (Magueijo et al 1995).

Therefore, for particularly small fields ($\la 2$ degrees), the
sampled power spectrum $C^s(k)$ is a bad estimator for the underlying CMBR
power spectrum $C(k)$, even after smoothing out field edge effects.  However,
in spite of the examples given, as the size of the field increases the sampled
power spectrum rapidly converges to the underlying power spectrum for $k >
k_{\rm min}$. In fact {\em so long as edge effects are dealt with}, for fields
as small as 10 degrees for a square window (with cosine bell applied) or 5
degrees FWHM for a Gaussian window, the sampled power spectrum $C^s(k)$
faithfully reflects the underlying Doppler peak structure (see for instance
Fig.~\ref{fig5}). For these fields, apart from the white-noise tail for 
$k < k_{\rm min}$, one then has
\[
{\sf C}^s(\bmath{k}) \approx \alpha C(k)|\widetilde{B}(\bmath{k})|^2,
\]
where the normalisation constant $\alpha$ is given by
\begin{equation}\label{cn}
\alpha=\int {d^2\bmath{k}\over (2\pi)^2} |\widetilde{W}(\bmath{k})|^2
=\int {d^2\bmath{x}\over (2\pi)^2} |W(\bmath{x})|^2
\equiv{\Omega_{s}\over (2\pi)^2}.
\end{equation}
Equation (\ref{cn}) defines the effective solid angle $\Omega_s$ of the 
sampled field. In the previous plots $C^s(k)$ was divided by $\alpha$ to
give the correct normalisation.

\subsection{Correlations in the sampled power spectrum}
\label{citsps}

As we have seen, finite fields have the effect of distorting the underlying
CMBR power spectrum by convolving it with the function
$|\widetilde{W}(\bmath{k})|^2$. Aside from smoothing the underlying power
spectrum, this convolution also has the effect of inducing correlations
between the values of the sampled power spectrum $C^s(k)$ for different values
of $k$. To investigate these correlations further, we must first consider the
correlations induced in the sampled Fourier modes $a_s(\bmath{k})$.

\begin{figure}
\centerline{\epsfig{
file=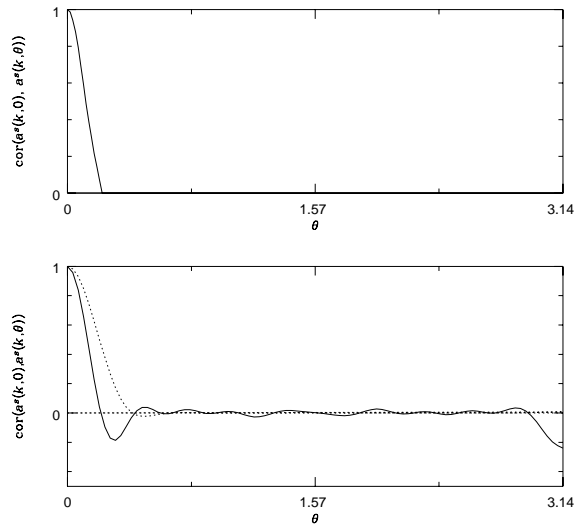,
width=8cm}}
\caption{The correlation of sampled Fourier modes 
${\rm cor}[a_s(50,0),a_s(50,\theta)]$ as a function of $\theta$: (top) for a
Gaussian window with $\sigma_w=12$ degrees; (bottom) for a square window with
$L=36$ degrees (solid line), and same field but multiplied by a cosine bell
(dashed line).  Correlations fall-off beyond $\theta\approx 1/\sigma_w$ or
$\theta\approx 1/L$, except for fields with sharp edges, where long distance
correlations persist.  In particular, for a square window there exist large
anticorrelation between modes symmetric about the origin of
$\bmath{k}$-space.}
\label{fig4}
\end{figure}

We expect correlations to exist between neighbouring sampled Fourier modes
within some correlation length $\xi$. However, if the window has sharp
edges, and the raw $C(k)$ spectrum falls-off as $1/k^2$ or slower at low $k$,
then in addition to nearby correlations there are also `resonant' long range
correlations even for relatively large fields.
We illustrate this point by considering the correlation of
sampled modes around a given circle in $\bmath{k}$-space, i.e.  ${\rm
cor}[a_s(k,0),a_s(k,\theta)]$ as a function of $\theta$ for some given radius
$k$. This quantity may be easily calculated from from (\ref{covmat}) or
(\ref{covmat2}), and is plotted in Fig.~\ref{fig4} for $k=50$ for
various windows.

For an inteferometer observation with a Gaussian window 
the correlations fall-off at a correlation length of order
$\xi\approx1/\sigma_w$ around each mode. Furthermore, there are no long-range
(anti-)correlations, and indeed no {\em anti}-correlations whatsoever. For
single-dish observations with a square window, however, apart
from correlations between neighbouring modes up to $\xi\approx 1/L$, there
also exist `resonant' anti-correlations between modes symmetric about the
origin. This effect only disappears for $k>k_b/2$, where $k_b$ defines the
angular resolution of the single-dish observation as in (\ref{ftbeam}).
This behaviour may be suppressed by multiplying the field by a cosine bell,
as discussed above, and the result is also shown in Fig.~\ref{fig4}.

The correlations between Fourier modes strongly affect power spectra 
estimation. Let us consider the {\em observed} sampled power spectrum
\begin{equation}\label{cskest}
C_s(k)={\int^{2\pi}_0}{d\theta\over 2\pi}|a_s(k,\theta)|^2,
\end{equation}
(which in keeping with our notation we write with the $s$ downstairs).
Although $\langle C_s(k)\rangle=C^s(k)$ (as expected), the
correlation of the Fourier modes induces correlations between $C_s(k)$ and
$C_s(k')$ for $|k-k'|<\xi$. These correlations are troublesome as they may
impart features on any observed spectrum $C_s(k)$ which average out to zero in
the ensemble average spectrum $C^s(k)$. This renders $C_s(k)$ a bad estimator
for $C^s(k)$, and conversely makes $C^s(k)$ a poor prediction for what an
observers actually sees. For example, as we show later, the low $k$ white
noise tail $C^s(k)=\lambda$ (for some constant $\lambda$) corresponds to a
{\em single} correlated piece of the spectrum.  As a result a typical observer
does not see a spectrum $C_s(k)\approx \lambda$ spectrum with small random
fluctuations around $\lambda$, but in general each observer sees a spectrum of
the form $C_s(k)\approx \mu\ne\lambda$ with small random fluctuations around
$\mu$. Although the ensemble average of $\mu$ equals $\lambda$, no single
observer could ever guess it.  In general even if the field size is large
enough so that $C^s(k)$ is a fair representation of 
the underlying CMBR power spectrum
$C(k)$, this does not mean
that any observed $C_s(k)$ would actually let us estimate $C^s(k)$. This,
so-called, cosmic covariance problem is well known in the context of
non-Gaussian theories, where it is present even if one assumes all-sky
coverage (Magueijo 1995). 

The best way to deal with cosmic covariance is to do
away with the correlations. We shall do this by discretizing the spectrum
estimates in such a way as to obtain a maximal finite set of uncorrelated
estimates. The separation between these uncorrelated estimates
will then define the spectral resolution $\Delta k (\approx \Delta\ell)$.

We could attempt to obtain uncorrelated estimates by using the estimator 
$C_s(k)$ as defined in (\ref{cskest}) and discretizing directly in 
the space $k=|\bmath{k}|$. The covariance of $C_s(k)$ at different
$k$ values is given by
\begin{equation}\label{covmark}
{\rm cov}[C_s(k),C_s(k')]=\int^{2\pi}_0 {d\theta\over 2\pi}
{\rm cov}[|a_s(k,0)|^2,|a_s(k',\theta)|^2],
\end{equation}
which may be simplified significantly if the window $W(\bmath{x})$
used in the experiment is real and 
even (as are all those considered here).  In this
case its Fourier transform $\widetilde{W}(\bmath{k})$ is real, and the real and
imaginary parts of the $a_s(\bmath{k})$ modes are uncorrelated.  For two
(possibly correlated) complex Gaussian random variables $z_1$ and $z_2$ with
uncorrelated real and imaginary parts, it can be shown that ${\rm
cov}[|z_1|^2,|z_2|^2]={\langle z_1z^*_2\rangle}^2+ {\langle z_1z_2\rangle}^2$,
and so
\[
{\rm cov}[|a_s(\bmath{k})|^2,|a_s(\bmath{k}')|^2]=
\langle a_s(\bmath{k})a_s^*(\bmath{k}')\rangle^2 + 
\langle a_s(\bmath{k})a_s(\bmath{k}')\rangle^2.
\]
Substituting this result into (\ref{covmark}) and using the fact that
the sky is real, we find
\begin{equation}
{\rm cov}[C_s(k),C_s(k')]=2\int^{2\pi}_0 {d\theta\over 2\pi}
\langle a_s(k,0)a_s^*(k',\theta)\rangle^2.  
\end{equation}
The integrand in this expression may be computed from (\ref{covmat}) or
(\ref{covmat2}), and it can be checked numerically that there is always a
significant correlation for $|k-k'|<\xi$, where $\xi$ is the correlation
length discussed above.  This can be understood from the fact that, roughly
speaking, each mode $a_s(\bmath{k})$ is correlated with neighbouring modes
within a circle of radius $\xi$ in the $\bmath{k}$-plane. Since the esitimator
$C_s(k)$ makes use of modes in all directions there are always correlations
between two estimates $C_s(k)$ and $C_s(k')$ for $|k-k'|<\xi$.

It is clear, however, that the separation $\Delta k$
between independent estimates of the sampled power spectrum should increase
with $k$ since the number of independent Fourier modes per unit $k$
increases. Therefore, instead of using the estimator $C_s(k)$ as defined in
(\ref{cskest}), and attempting to remove correlations by discretizing directly
in $k=|\bmath{k}|$ space, we should first discretize the modes in the 
two-dimensional $\bmath{k}$ space.

\subsection{Uncorrelated meshes and the spectral resolution $\Delta l$}
\label{umasr}

From equations (\ref{css}) and (\ref{csi}) we see that the presence of the 
window $W(\bmath{x})$ causes the underlying (two-dimensional) CMBR
power spectrum to be convolved with $|\widetilde{W}(\bmath{k})|^2$. To obtain
independent estimates of the sampled power spectrum, we may divide the 
$\bmath{k}$-plane into a set of discrete cells so that for the points
at the centres of any two cells (or infact any points which have the
same relative position in each cell) we have
\begin{equation}
{\rm cor}[a_s(\bmath{k}_i),a_s^*(\bmath{k}_j)]< \epsilon,
\label{cellcor}
\end{equation}
where $\bmath{k}_i$ and $\bmath{k}_j$ are the `centres' of the $i$th
and $j$th cell respectively, and the constant $\epsilon \ll 1$.  More will be
said about $\epsilon$ and how small it must be in the next section. Although
we may divide the $\bmath{k}$-plane in any way we wish, so long as
(\ref{cellcor}) is satisfied, we choose to discretize in the form of a
rectangular mesh.  Therefore working outwards along the axis corresponding to
the smallest dimension of the field, we find the set of points for which ${\rm
cor}[a_s(\bmath{k}_i),a_s(\bmath{k}_{i+1})]=\epsilon$, and then repeat the
process for the perpendicular axis. These point then define our rectangular
mesh. Normally the mesh will not be very different from a square lattice with
cell size $k_0\approx 2\pi/{\sqrt \Omega_s}$, where $\Omega_s$ is the
effective solid angle of the field as defined in (\ref{cn}).

From Fig.~\ref{fig4} we see that such a
procedure is impossible for, say, a square window, because of the long range
correlations it induces. In such a case a the field must first be multiplied
by the cosine bell (\ref{bell}) before the uncorrelated mesh can be 
constructed.  Once the uncorrelated mesh has been 
built we may use a new power spectrum estimator
\begin{equation}
C_s(k)={1\over N(k)}{\sum_{|\bmath{k}_i|=k}} |a_s(\bmath{k}_i)|^2
\label{newest}
\end{equation}
where the sum is over all modes in the mesh for which $|\bmath{k}_i|=k$,
and $N(k)$ is the number of such modes. From (\ref{newest}) we see that
the sampled power spectrum is estimated only at a finite number of well 
separated values of $k$, and that these estimates are 
quasi-uncorrelated. Moreover the estimates are distributed as
a $\chi^2_{N(k)}$ distribution. 
For a square lattice in which the modes are separated by $k_0$, the
values of $k$ for which the sampled power spectrum is estimated
are those where $k=k_0{\sqrt {n_1^2+n_2^2}}$ for integer numbers $n_1$
and $n_2$. The number of modes at a given $k$ is simply the number
distinct permutations of integers $n_1$ and $n_2$ for which
$\sqrt{n_1+n_2}$ is fixed. For instance, for $k=k_0$ we have $N(k)=4$, whereas
for $k=5k_0$ we find $N(k)=8$. 
It is only for $k\gg (k_0^2/2\pi)$ that the number of 
independent modes between
$k$ and $k+1$ becomes much larger than 1 
(since for large $k$ the number of such modes is $\approx2\pi k/k_0^2$). 
From then on a resolution $\Delta \ell \approx \Delta k=1$
is meaningful and $C_s(k)$ should be replaced  by
\begin{equation}
C_s(\ell)={1\over N(\ell)}{\sum_{\ell<|\bmath{k}_i|<\ell+1}}  
|a_s(\bmath{k}_i)|^2   
\label{newest2}
\end{equation}
where $N(\ell)$ is the number of mesh points for which 
$\ell<|\bmath{k}_i|<\ell+1$. 

In practice it is the fact that we can only make uncorrelated estimates with a
separation $\Delta k$ that limits the spectral resolution.  Spectral leakage
also constrains $\Delta k$ but in the context of Doppler peak detection we
will see that whenever there are enough independent modes to resolve the
peaks, spectral leakage is no longer a problem. Also in that regime the
underlying $C(k)$ spectrum does not change by much across each of the cells of
the uncorrelated mesh. Hence each of the independent estimates $C_s(k)$
provided by the mesh for $k< (k_0^2/2\pi)$ could be placed anywhere in between
its two neighbours (but of course only at one such point at a time). 

Spectral resolution understood in this sense leads to the following picture
for an experiment with a sky coverage area $\Omega_s$ (for which $k_0\approx
2\pi/{\sqrt \Omega_s}$).  The region in the spectrum between $k=0$ and 
$k=k_0$
receives only one independent estimate, which bears little resemblance to the
true underlying spectrum.  For $k \ga k_0$ the sampled spectrum and the raw
spectrum are roughly proportional, and the average separation $\Delta k$
between uncorrelated estimates goes like $\Delta k\approx {\sqrt
{k^2+k_0/\pi}}-k$.  For $k\gg k_0$ this means $\Delta k\approx k_0^2/(2\pi
k)$, so that the spectral resolution goes as $1/k$. A meaningful resolution
of $\Delta \ell \approx \Delta k=1$ can only be achieved for 
$k\gg k_0^2/(2\pi)$.  For example, if an experiment has a field of size
10 degrees, then  $k_0\approx 36$, and so we expect
$C^s(k)\propto C(k)$ for $k \ga 60$. 
However, one may not achieve independent estimates of the spectrum
with separation $\Delta l\approx\Delta k=1$ before $k\gg 250$.  
Uncorrelated meshes provide
the maximal spectral resolution $\Delta l$ for a given experiment. Of course
these uncorrealted estimates may be grouped into wider bins, which
deliberately reduces the spectral resolution, but also reduces the 
cosmic/sample variance in each of the estimates.

\section{Cosmic variance in small fields}\label{cv}

We now compute the cosmic/sample variance in the estimates provided by
uncorrelated meshes both with and without instrumental noise.
Once these are computed, the cosmic/sample variance in estimates obtain by any
other binning of uncorrelated estimates (as the one proposed in
Section~\ref{peakstat}) is straightforward.

\subsection{Variance of estimators without noise}

Ignoring instrumental noise a centred estimator of the underlying CMBR
power spectrum $C(k)$ is given by
\[
{\overline C}(k)={C_s(k)\over \alpha |\widetilde{B}(k)|^2}=
{1\over N(k)\alpha |\widetilde{B}(k)|^2}
{\sum_{|\bmath{k}_i|=k}} |a_s(\bmath{k}_i)|^2,
\]
where $\alpha=\Omega_s/(2\pi)^2$ as defined in equation (\ref{cn}). It is 
straightforward to verify that indeed ${\langle{\overline C}(k)\rangle}=C(k)$,
and by performing a calculation similar to that following equation 
(\ref{covmark}), we find the variance of this estimator to be given by
\begin{equation}\label{csvar}
\sigma^2({\overline C}(k))={2C^2(k)\over \widetilde{N(k)}},
\end{equation}
where $\widetilde{N}(k)$ is given by
\begin{equation}
\widetilde{N(k)}=N(k)^2 /{\sum_{|\bmath{k}_i|=k \atop |\bmath{k}_j|=k}}
\{{\rm cor}[a_s(\bmath{k}_i),a_s^*(\bmath{k}_j)]\}^2 
\end{equation}
The quantity $\widetilde{N_k}$ acts as an effective number of independent modes
contributing to the estimate ${\overline C}(k)$. If
all the $N(k)$ modes were perfectly correlated or anticorrelated,
then they would only count as one, and in that case $\widetilde{N(k)}=1$. 
If, on the other hand,
all the $N(k)$ modes are very nearly uncorrelated, then 
$\widetilde{N(k)}\approx N(k)$. Following a similar calculation to the above
the correlations between two estimates can is found to be
\[
{\rm cor}[{\overline C}(k),{\overline C}(k')]=
{2\over N(k)N(k')}\sum_{|\bmath{k}_i|=k \atop |\bmath{k}_j|=k'}
\{{\rm cor}[a_s(\bmath{k}_i),a_s^*(\bmath{k}_j)]\}^2.
\]
If an `uncorrelated' mesh is built such that for any two mesh points
\[
{\rm cor}[a_s(\bmath{k}_i),a_s^*(\bmath{k}_j)]< \epsilon,
\]
as discussed in section \ref{umasr}, then
residual correlations of order $\epsilon^2/\sqrt{N(k)N(k')}$
will persist between the estimates ${\overline C}(k)$ 
and ${\overline C}(k')$.
In this case we also find $\widetilde{N(k)}\approx N(k)(1-\epsilon^2)$.
 
These results allow a more concrete definition of the uncorrelated mesh. The
point of uncorrelated meshes is to do away with correlations among estimates
without throwing away information. However reducing the cosmic/sample variance
in an estimate and reducing the correlations between estimates are
contradictory requirements.  In other words, by lowering the value of
$\epsilon$ arbitrarily, one reduces correlations between the ${\overline
C}(k)$ estimates, but their individual variances increase.

The best compromise is obtained by considering the 
effective number of independent modes {\em within} a mesh cell
centred on $\bmath{k}_i$, which is given by
\[
\widetilde{N}_{\rm cell}(\bmath{k}_i)=
1/ \int_{\rm cell}{d^2\bmath{k}\over A_{\rm cell}}
\{{\rm cor}[a_s(\bmath{k}_i),a_s^*(\bmath{k})]\}^2,
\]
where $A_{\rm cell}$ is the area of the mesh cell. Clearly, 
$\widetilde{N}_{\rm cell}(\bmath{k})$ is always greater than unity. 
One can therefore, on average, avoid the 
loss of non-redundant information by computing the average
density of independent modes around a given mode at $\bmath{k}=\bmath{k}_i$
\[
\rho(\bmath{k}_i)=1/ \int d^2\bmath{k}
\{{\rm cor}[a_s(\bmath{k}_i),a_s^*(\bmath{k})]\}^2,
\]
and defining the mesh size as $k_0=1/{\sqrt{\rho}}$. In this way
power spectrum estimators derived from large regions of the $\bmath{k}$-plane
have the same variance whether one uses the mesh
or the continuum of modes in its calculation.

In well behaved regions where $C^s(k)\propto C(k)$ we find that that
$k_0={\sqrt {2\pi}}/\sigma_w$ for a Gaussian window, and (of course)
$k_0=2\pi/L$ encodes all the information in a square field (multiplied by a
cosine bell). This choice of $k_0$ corresponds to $\epsilon\approx 0.2$. We
may may check that even such a large value of
$\epsilon$ is small enough to make all
correlations negligible.  Since in the estimator ${\overline C}(k)$ no two
{\em neighbouring} cells are ever used for the same $k$, one may safely set
$\widetilde{N}(k) \approx N(k)$ when computing $\sigma^2({\overline C}(k))$.
Furthermore the residual correlations between ${\overline C}(k)$ and
${\overline C}(k')$ involves at most two neighbouring modes, and so are of the
order $0.04/N(k)$, which is typically smaller than 1 per cent. Hence one need
not throw away any information in order to keep residual correlations low, as
uncorrelated meshes spread the modes among estimates in the best possible way.

For $k\gg (k_0^2/2\pi)$ individual $C^{\ell}$ coefficients may be estimated by
\[
{\overline C}(\ell)={C_s(\ell)\over \alpha|\widetilde{B}(\ell)|^2}=
{1\over \alpha |\widetilde{B}(\ell)|^2N(\ell)}
{\sum_{\ell<|\bmath{k}_i|<\ell+1}}  |a_s(\bmath{k}_i)|^2.
\]
For a square window of side $L$ the variance in this estimator is given
by
\begin{equation}
\sigma^2[{\overline C}(\ell)]={2C^2(\ell)\over \widetilde{N}(\ell)}
\approx {C^2(\ell)\over \ell}{4\pi\over L^2},
\label{sigl}
\end{equation}
since 
$\widetilde{N}(\ell)\approx N(\ell)\approx 2\pi \ell/k_0^2\approx L^2l/2\pi$.
Equation (\ref{sigl}) has a simple interpretation.  The cosmic variance of
$C(\ell)$ is simply $2C^2(\ell)/(2\ell+1)\approx C^2(\ell)/\ell$ for large
$\ell$, and $4\pi/L^2$ is the fraction of the sky covered by the experiment.
However, this formula is only a good approximation for $\ell \approx k\gg
(k_0^2/2\pi)$, and significant corrections are necessary for smaller values of
$\ell$.

Finally, we note that by increasing the size of the observed field one does
not decrease the relative error in the estimate provided by each of the modes
$a_s(\bmath{k})$.  The relative error in each of the modes is fixed by its
Gaussian nature, and its value $\sigma^2(|a_s(\bmath{k})|^2)=2C^{s2}(k)$
merely reflects the vanishing kurtosis of a Gaussian distribution.  The
positive effect of increasing the size of the field is that it populates
Fourier space more densely with uncorrelated modes.  This improves the
spectral resolution and also reduces the cosmic variance in the estimates
based on a larger number of independent modes.

\subsection{Observations of multiple fields}

So far we have assumed that observations are of a single finite field.  It may
sometimes happen, however, that the total set of observations consists of
$n_f>1$ fields each observed with a window $W_i({\bf x})$ for $i=1$ to $n_f$.
If $a_s(\bmath{k})$ and $b_s(\bmath{k})$ are the sampled modes derived from
observations of the $i$th and $j$th field respectively, then it is
straightforward to show that their covariance matrix is given by
\begin{equation}
\langle a_s(\bmath{k})b_s^*(\bmath{k}')\rangle=
\int {{d^2\bmath{k}''}\over (2\pi)^2} C(\bmath{k}'')
\widetilde{W}_i(\bmath{k}-\bmath{k}'')\widetilde{W}_j(\bmath{k'}-\bmath{k}'').
\end{equation}
If the $j$th window can be obtained from the $i$th window by a translation
with vector $\bmath{R}$ then 
\[
\widetilde{W}_j(\bmath{k})=W_i(\bmath{k})\exp(i\bmath{R}\cdot\bmath{k}),
\]
and the only chance for a correlation is therefore if 
$\bmath{k}=\bmath{k}'$. However, whenever the approximation 
$C^s(k)\propto C(k)$ is valid, we find that
\begin{eqnarray*}
{\rm cor}[a_s(\bmath{k}),b_s^*(\bmath{k})] 
& \approx &
{{\int {d^2\bmath{k}\over (2\pi)^2}|\widetilde{W}(\bmath{k})|^2
\exp(i\bmath{k}\cdot\bmath{R})}
\over {\int {d^2\bmath{k}\over (2\pi)^2}|\widetilde{W}(\bmath{k})|^2}} \\
& \approx &
{{\int d^2\bmath{x}~W(\bmath{x})W^*(\bmath{x}+\bmath{R})}
\over {\int d^2 \bmath{x} |W(\bmath{x})|^2}},
\end{eqnarray*}
which is of the order of the percentage of area overlap between the windows.
Therefore, if the fields are well-separated then
correlations between the sampled modes from each field are very small. 
In such a case one can
build an uncorrelated mesh for each field, 
and superpose the meshes giving an extra index
$i=1,\ldots, n_f$ to each of the mesh points. The uncorrelated 
mesh estimates in well behaved regions [where $C^s(k)\propto C(k)$]
will then have the same properties as the single field estimates but with
$N(k)$ replaced by $n_fN(k)$. Therefore the relative error in the 
uncorrelated estimates
of the sampled power spectrum {\em can} be reduced by observing several
well-separated fields, or (for single dish observations) 
by dividing observations of a single large field into several smaller
(but non-overlapping) fields.

\section{Estimating the CMBR power spectrum from real observations}
\label{noise}

So far we have only considered the effects of the window and the beam
of an experiment on the sampled power spectrum. Real observations, however,
have the added difficulties of instrumental noise, 
the discretization of the sky temperature distribution into pixels
(for single-dish experiments), and the 
existence of foreground emission.

\subsection{The effects of noise and pixelisation}

We first  consider the effects of noise
and pixelisation on sampled power spectrum estimates drawn
from uncorrelated meshes. In our analysis we neglect residual correlations 
as well as any terms of order $\epsilon^2$.

Let us first consider single-dish observations, in which the CMBR sky
map of the field is discretized into $M$ pixels at positions $\bmath{x}_i$ and
with noise $N_i$ for $i=1$ to $M$. For simplicity we shall assume that the
$N_i$ are independent Gaussian random variables with $\langle N_i \rangle=0$
and
\[
\langle N_iN_j \rangle = \sigma_i^2 \delta_{ij}.
\]
This assumption is a reasonable approximation to reality. To simplify
our discussion still further we also assume that the variance of the noise in
each pixel is constant so that $\sigma_i=\sigma_{\rm pix}$. 
We can eliminate the 
discreteness by instead considering the noise as a continuous Gaussian random
field $N(\bmath{x})$ which is added to the sky temperature distribution, and
is characterised by the two-point correlation function
\begin{equation}\label{noise1}
\langle N(\bmath{x})N(\bmath{x}') \rangle 
\approx \delta (\bmath{x}-\bmath{x}')\sigma^2_{\rm pix}\Omega_{\rm pix}
\end{equation}
where $\Omega_{\rm pix}$ is the area of a pixel. Assuming all-sky coverage
these two quantities are often combined into $w^{-1}=\sigma^2_{\rm
pix}\Omega_{\rm pix}$, where $w$ may be considered as a weight per unit solid
angle (Knox 1995).  If $t_{\rm tot}$ is the total observational time
available, then the time spent observing any given pixel is $t_{\rm
pix}=t_{\rm tot}\Omega_{\rm pix}/(4\pi)$.  The noise variance per pixel
$\sigma^2_{\rm pix}=s^2/t_{\rm pix}$, where $s$ is the sensitivity of the
detector. Therefore, for a detector with fixed sensitivity $s$ and for a given
total observation time $t_{tot}$, by changing $\Omega_{\rm pix}$ the quantity
$w^{-1}=\sigma^2_{\rm pix}\Omega_{\rm pix}=4\pi s^2/t_{tot}$ remains
constant. Hence
$w^{-1}$ is important as a qualifier for noise in all-sky maps obtained using
different scanning strategies.  If we consider the most general
case where we make $n_{\rm f}$ maps each of size $\Omega_s$,
then the time spent on each pixel is  
$t_{\rm pix}=t_{\rm tot}\Omega_{\rm pix}/(n_{\rm f}\Omega_s)$, and the
the quantity which now remains constant is
\begin{equation}
w^{-1}
=\frac{4\pi s^2}{t_{\rm tot}}
=\sigma^2_{\rm pix}\Omega_{\rm pix}
\left(\frac{4\pi}{n_{\rm f}\Omega_s}\right).
\label{wsd}
\end{equation}

The noise defined by (\ref{noise1}) adds an extra term in the
covariance matrix of the sampled modes
${\langle a_s({\bf k}),a^{s*}({\bf k'})\rangle}$ of the form
\begin{equation}
\langle a_s^N(\bmath{k})a_s^{N*}(\bmath{k}')\rangle
={\sigma^2_{\rm pix}\Omega_{\rm pix}\over (2\pi)^2}
\int d^2\bmath{x} |W(\bmath{x})|^2e^{-i(\bmath{k}-\bmath{k}')\cdot\bmath{x}}.
\label{sdnoise}
\end{equation}
Since an actual sky map is divided into pixels of size $d$ (say), we would 
estimate the
sampled Fourier modes $a_s(\bmath{k})$ using a Fast Fourier Transform
(FFT). We would therefore obtain estimates of the Fourier modes 
between $k=0$ and $k=2\pi/d$ on a square grid with spacing $2\pi/L$, where
$L$ is the size of the field, i.e. one estimate per uncorrelated mesh cell.
Therefore, neglecting terms of order $\epsilon^2$ in (\ref{sdnoise}), we
find that the noise on the uncorrelated mesh points is approximately
\begin{equation}
\langle a_s^N(\bmath{k}_i)a_s^{N*}(\bmath{k}_j)\rangle=
\delta_{ij}\alpha \sigma^2_{\rm pix}\Omega_{\rm pix},
\label{sdnoise2}
\end{equation}
where $\alpha=\Omega_s/(2\pi)^2$ as defined earlier.

For interferometers, on the other hand, instrumental noise is added in Fourier
space directly, and may be quantified by means of the noise per mesh cell
$\sigma^2_N$, such that
\begin{equation}
\langle a_s^N(\bmath{k}_i)a_s^{N*}(\bmath{k}_j)\rangle
=\delta_{ij}\sigma^2_N
\end{equation}
We shall assume that the interferometer possesses sufficient antennas
appropriately positioned to provide roughly uniform coverage of the Fourier
domain in the $k$-range of interest.  A method for positioning a given number
of antennas in such a way is given by Keto (1996). It may, of course, be
possible to tailor the coverage of the Fourier plane so the density of
measured visibilities is greatest in the regions of the (one-dimensional)
$k$-range which are of most interest (so that the interferometer may be
thought of as a matched filter to the expected CMBR power spectrum).  This
does, however, rather preempt the results of the observation, and furthermore,
in addition to power spectrum estimation, interferometer experiments are
usually designed to make maps of the CMBR, for which uniform coverage is
desirable.

If the density of measured
visibilities (i.e. the number of visibilities per unit area in the
$\bmath{k}$-plane) is $\rho_{\rm vis}$, then the number of visibilities
per mesh cell is simply
\[
n_{\rm vis}=\rho_{\rm vis} k_0^2 
\approx \rho_{\rm vis} \frac{2\pi^2}{\Omega_s},
\]
where $k_0$ is the cell size. The measured visibilities in each cell
(say at $\bmath{k}=\bmath{k}_i$) may then be used to estimate
$|a_s(\bmath{k}_i)|^2$ by, for example, a maximum likelihood analysis
(see Hobson, Magueijo \& Kaiser, in preparation), or by some
weighted average in the limit of low noise. In any case, a lower limit
to the noise per mesh cell is given by
\begin{equation}
\sigma_N^2
=\frac{s^2\Omega_s^2}{n_{\rm vis}t_{\rm vis}}
=\frac{s^2\Omega_s^3}{2\pi^2\rho_{\rm vis}t_{\rm vis}},
\label{intnoise}
\end{equation}
where $t_{\rm vis}$ is the time spent measuring each visibility. The
presence of $\Omega_s^2$ (the area of the observed field) in the first
equality results from defining the Gaussian window of the experiment
to be always equal to unity at its peak. By changing the spacing of
the inteferometer antennas we may fix the limits $k_{\rm min}$ and
$k_{\rm max}$ of the $k$-range to which the experiment is
sensitive. In targetting differences between the
cosmological theories, or determining the existence (or otherwise) of
Doppler peaks, these limits are usually fixed. In any case, the 
the area of the
$\bmath{k}$-plane sampled is $A= \pi(k_{\rm max}^2-k_{\rm min}^2)$.
Now the time spent observing a particular field is given by
$t_{\rm f}=A\rho_{\rm vis}t_{\rm vis}$, but the total observing time
$t_{\rm tot}$ is divided among $n_{\rm f}$ fields then the time per
field is simply $t_{\rm f}=t_{\rm tot}/n_{\rm f}$.  Therefore for
comparing inteferometer observations with the same detector
sensitivity $s^2$, total observing time $t_{\rm tot}$ and
$\bmath{k}$-plane coverage $A$, but in which the size of the field
$\Omega_{\rm s}$ and the number of such fields $n_{\rm f}$ can vary,
then the quantity to keep constant is
\begin{equation}
w^{-1}=\frac{As^2}{t_{\rm tot}}=
\frac{2\pi^2\sigma_N^2}{n_f\Omega_s^3}.
\label{wint}
\end{equation}
We may then describe the noise per mesh cell by
\[
\langle a_s^N(\bmath{k}_i)a_s^{N*}(\bmath{k}_j)\rangle
=\delta_{ij}\frac{w^{-1}n_{\rm f}\Omega_s^3}{(2\pi)^2}.
\]

Noise always introduces a bias in the estimator
${\overline C}(k)$. Hence in the presence of noise one should
use instead the centred estimator for single-dish observations 
\begin{equation}
{\overline C}(k)=
\left({1\over n_{\rm f}N(k)\alpha |B(k)|^2}
\sum_{|\bmath{k}_i|=k} |a_s(\bmath{k}_i)|^2\right)
-{\sigma^2_{\rm pix}\Omega_{\rm pix}\over |B(k)|^2}
\end{equation}
and for interferometers
\begin{equation}
{\overline C}(k)={\left(
{1\over n_{\rm f}N(k)\alpha}{\sum_{|\bmath{k}_i|=k}}  |a_s(\bmath{k}_i)|^2
\right)} -{\sigma^2_N\over \alpha}
\end{equation}
For $k\gg (k_0^2/2\pi)$ these estimators
should be rewritten in the usual way. The variances in these
estimators are increased by the presence of noise. For single-dish
experiments
\begin{eqnarray}\label{varnoise1}
\sigma^2({\overline C}(k))\over C^2(k)
& = &
{2\over n_{\rm f}N(k)}\left(1+
{\sigma^2_{\rm pix}\Omega_{\rm pix}\over |B(k)|^2 C(k)}\right)^2\nonumber \\
& = &
{2\over n_{\rm f}N(k)}\left(1+
{w^{-1}\Omega_s n_{\rm f}\over4\pi |B(k)|^2 C(k)}\right)^2,
\end{eqnarray}
whereas for interferometers
\begin{eqnarray}\label{varnoise2}
\sigma^2({\overline C}(k))\over C^2(k)
& = &
{2\over n_{\rm f}N(k)}
\left(1+{\sigma^2_N \over \alpha C(k)}\right)^2 \nonumber\\
& = & 
{2\over n_{\rm f}N(k)}
\left(1+{w^{-1}\Omega_s^2 n_{\rm f}\over C(k)}\right)^2.
\end{eqnarray}

\subsection{The effects of smooth foreground emission}

Depending on the frequency at which the observations are made, the
existence of various types of foreground emission can severely hamper
the measurement of CMBR anisotropies.  A discussion of these
foreground components, and the regions of frequency/multipole space in
which each dominates, is given by Tegmark \& Efstathiou (1996) (TE96).  The
main components of this foreground emission are extragalactic radio
point sources, and continuous Galactic dust, 
synchrotron and free-free emission.

The general method for dealing with continuous 
Galactic foreground emission is to
make observations at several different frequencies, and use this spectral
information (together with their predicted power spectra in 
the Wiener filtering approach as in TE96) 
to perform a subtraction of these components. This process
results in estimates of temperature maps or Fourier modes due to the
CMBR {\em alone}, together with some errors on these estimates. 

In general the power spectrum of the these separation errors is not
strictly constant across the $\bmath{k}$-plane, but may often be
approximated as such. We may then model the reconstruction errors by
the presence of a generalised noise field, and include the effects of
foreground subtraction simply as a factor which multiplies the noise
per mesh cell. Therefore, in the expression (for single-dish
observations) for $w^{-1}$ (\ref{wsd}), the quantities $\sigma_{\rm
pix}$ and $\Omega_{\rm pix}$ refer to the pixel noise and pixel
solid-angle in the {\em final} CMBR temperature map deduced from
observations at several frequencies using some separation algorithm.
Similarly, $\sigma_N$ in (\ref{wint}) refers to the error on the {\em
deduced} CMBR Fourier modes.

Clearly the size of the errors in the CMBR map or power spectrum will
depend on the value of $w^{-1}$ for each frequency channel, on the
number of such channels, and on the separation algorithm used.  TE96
show that for a `pixel-by-pixel' subtraction algorithms these errors are
typically several times larger than the average error on an individual
frequency channel due to instrumental noise alone. Using a vector Wiener
filtering algorithm, however, these errors can be reduced by up to a
factor of 10, depending on the number of frequency channels and their
range in frequency. Nevertheless, as the authors themselves suggest, the
Wiener filtering method is rather optimistic in that it
assumes that both the frequency dependence and the power spectra
of the various foreground components are known reasonably well, which
is certainly not the case. A reasonable estimate might be that
the errors associated with the separation process are of a similar
magnitude to the average errors on an individual frequency channel due to
instrumental noise alone.

Finally, we note that since we are using an uncorrelated mesh in the
estimation of the CMBR power spectrum, the additional errors on the
power spectrum estimates arising from any foreground separation
process will still be uncorrelated.

\subsection{The effects of point sources}\label{point}
 
As mentioned above the existence of foreground point sources can cause
problems for estimating the CMBR power spectrum. In general point
sources cannot be removed from spectral information alone, but require
the identification of the sources by higher-resolution observations.
The general scheme is to survey the same region of sky as that
observed by the CMBR experiment at a frequency close to that of the
CMBR observations in order to indentify all the point sources down to
some flux limit, which are then subtracted from the CMBR data.  The
required flux limit is usually such that the confusion noise from
unsubtracted sources is roughly equal to the instrumental noise. Once
this source-subtraction has been performed the residual emission from
unsubtracted sources can be modelled as a an additional smooth
foreground component, and treated in the same way as the Galactic
emission, as discussed above.

We note here that although it is generally believed that point source
contamination becomes less important as the observing frequency
increases above about 100 GHz, there is no direct evidence for
this. Moreover, even the population of radio point sources at
frequencies above about 10 GHz is rather uncertain, and it may be
inadvisable to rely on low frequency surveys such as the 1.5GHz VLA
FIRST survey (Becker et al. 1995) to subtract point sources from CMBR
maps made at much higher frequencies.  In order that one can be
confident in the final CMBR map/power spectrum, it is therefore
necessary to make higher-resolution observations at frequencies close
to those of the CMB experiment. Therefore, if the CMB observations are
made over a large frequency range, higher-resolution observations at
{\em several} frequencies in this range may be required.  Clearly,
this can {\em severely} limit the maximum possible sky coverage, and
this should be borne in mind when interpreting the figures presented
in the next section.  The detailed implications of source subtraction
for total sky coverage will be discussed fully in a forthcoming paper.

\section{Observing secondary peaks of standard CDM}\label{peakstat}

\begin{figure}
\centerline{\epsfig{
file=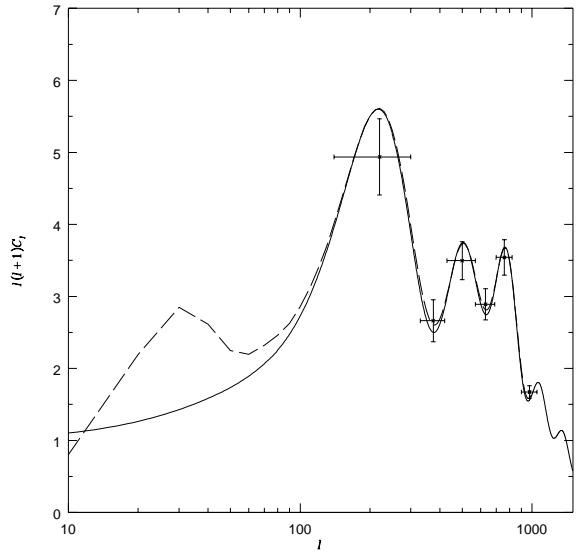,
width=8cm}}
\caption{sCDM (line) as seen by a square window with $L=10$ degrees after
and multiplication by a cosine bell (dash). The points indicate 
the corresponding power spectrum estimates for a particular realisation
in each of the bins indicated by the horizontal error bars. The vertical
error bars indicate the theoretical cosmic/sample in the absence of
noise. We have set $B(k)=1$ for clarity.}
\label{fig5}
\end{figure}

If one is interested only in determining whether or not
secondary oscillations exist in the CMBR power spectrum, then
the uncorrelated mesh estimates should be combined into broader bins,
centred on the anticipated positions of the peaks and troughs in the
power spectrum under consideration. Six broad bins
adjusted for testing standard CDM (sCDM) are shown in Fig.~\ref{fig5}.  
The figure shows the ensemble average sampled power spectrum for an
experiment with a square window with $L=10$ degrees (dashed line),
multiplied by a cosine bell, as compared to the underlying power
spectrum (solid line).  The points indicate the average power in each
of the bins denoted by the the horizontal errorbars. These points are
calculated by averaging the values of ${\overline C}(k)$ at each
measured $k$-value in each bin.  The vertical errorbars are
due to cosmic/sample variance on these estimates (assuming no noise).
They were computed by adding the variances of the measured
${\overline C}(k)$ estimate values in each bin, and dividing by the
square of the number of such estimates in the usual manner.

Once the bins have been chosen a statistic for the presence of
oscillations in the CMBR power spectrum can be derived by inferring
the convexity of the power spectrum at each bin position (apart
from the first and last bins).  If the estimated power in the $i$th
bin in $C_i$ (where here $i=1$ to 6), then we define (for $i=1$ to 4)
the quantities
\begin{equation}
{\cal C}_i={C_{i-1}+C_{i+1}\over 2}-C_i.
\end{equation}
These `convexities' are all negative if there are no
secondary peaks, but alternate in sign for sCDM. 
The variance in the $i$th convexity is simply
\begin{equation}
\sigma^2({\cal C}_i)=\sigma^2(C_i) + {\sigma^2(C_{i-1})+
\sigma^2(C_{i+1})\over 4}.
\end{equation}
Therefore a convexity in bins 2 or 4
which is larger than zero by an amount $n\times\sigma({\cal C}_i)$
corresponds  to a $n$-sigma detection of secondary oscillations.
Hence we can define an oscillation detection function as 
\begin{equation}
{\Sigma}_i={|{\langle {\cal C}_i\rangle}|\over \sigma
({\cal C}_i)}
\end{equation}
for the relevant $i=2$ and $i=4$. The function ${\Sigma}_i$ tells us
to within how many sigmas we can claim a detection of secondary peaks.
More generally we could define in a similar way a set of oscillation
statistics centered at a generic point $\ell=\ell_i$ in the power
spectrum and making use of bins with width $\Delta\ell$. These
statistics would then measure the oscillatory character of the
spectrum as a function of the point $l_i$ and the scale $\Delta\ell$.
In what follows, however, we shall always target particular theories,
and so we will not use this broader class of statistics.

\begin{figure}
\centerline{\epsfig{
file=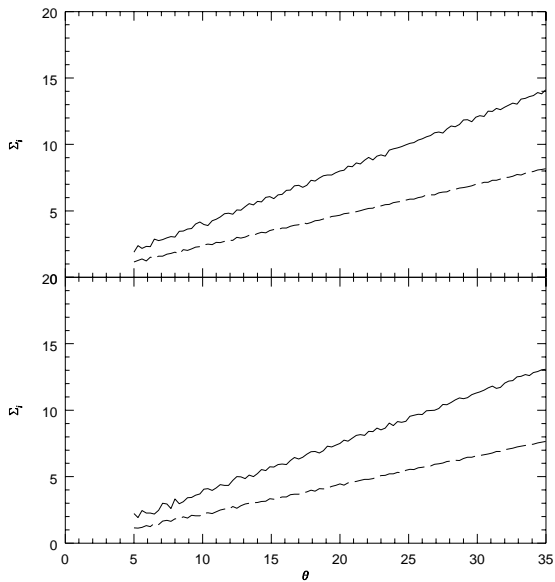,
width=8cm}}
\caption{${\Sigma}_2$ (line) and ${\Sigma}_4$ (dash) functions
for interferometers (top) and single-dish experiments (bottom).
We have assumed no noise ($w^{-1}=0$) and 
infinite resolution ($\sigma_b=0$). For the top plot $\theta$ is the
FWHM of the interferometer primary beam in degrees 
(FWHM $\approx 2.35\sigma_w$). For the
bottom plot $\theta$ is the side $L$ of a square field in degrees.}
\label{fig6}
\end{figure}

In Table~\ref{table} we display the results of applying the convexity
statistic to sCDM under the experimental situation used in
Fig.~\ref{fig5}, and we see that a very significant detection can be
obtained for this relatively small field ($L=10$ degrees) in the
idealized case of no noise ($w^{-1}=0$) and infinite resolution
($\sigma_b=0$).  
\begin{table}
\begin{center}
\leavevmode
\begin{tabular}{|c|c||c|c|c|c|} \hline
$\ell_1$  & $\ell_2$ &  $C_i$ & $\sigma(C_i)$ & ${\cal C}_i$ & 
${\Sigma}_i$\\ \hline
140 & 300 & 5.21 &0.78&      &      \\
330 &  420 & 2.69& 0.54&1.67&   3.75 \\
430 &570& 3.53&0.51&  -0.73  &   2.25  \\
570 & 690 & 2.89 &0.46&0.64  &   2.27     \\
700 & 820 & 3.54 &0.50 &-1.26  &    4.61    \\
900 & 1050 &1.67&0.30& & \\ \hline
\end{tabular}
\end{center}
\caption{Bins $(\ell_1,\ell_2)$ used if Fig.~\ref{fig5}, the average power
in bin $C_i$ and its variance, and convexities ${\cal C}_i$
together with the number of sigmas ${\Sigma}_i$
within which one could be certain of observing their correct sign.}
\label{table}
\end{table}
In Fig.~\ref{fig6} we plot the dependence of the
detection functions $\Sigma_2$ and $\Sigma_4$ on sky coverage for the
same idealized case. From the figure it is clear that a total sky
coverage of greater than about (5 deg)$^2$ is necessary for a
detection of secondary peaks. This supports our earlier claim that the
convolution of the power spectrum is never a practical problem in this
context (once the field edges-if they exist- have been smoothed by a
multiplication with a cosine bell).  The deterioration of the
detection function for fields smaller than this is, however,
considerably worse than simple extrapolation of our curves in
Fig.~\ref{fig6} imply.  From Fig.~\ref{fig6} we also see that 
the first dip in the sCDM power spectrum is more
easily detected than than the second one, a situation only exarcebated
by finite resolution and the prescence of instrumental noise.
Therefore, in the rest of this section, we shall confine ourselves to
considering the detection function $\Sigma_2$, which from now on we
refer to simply as $\Sigma$.

We now undertake a general analysis of the $\Sigma$ function
for single-dish and interferometer experiments separately, taking into account
the effects of instrumental noise and finite resolution.

\subsection{Single-dish experiments}

For single-dish observations the detection function
$\Sigma=\Sigma(w^{-1},\theta_b,L)$, i.e. it is a function of the
noise level $w^{-1}$ (which itself depends on the total observation
time and detector sensitivity), 
the beamsize characterised by $\theta_b\approx 2.35\sigma_b$, and the size of
the observed field $L$. We now consider two sections through this 
(three-dimensional) function for fixed $L$. 

\begin{figure}
\centerline{\epsfig{
file=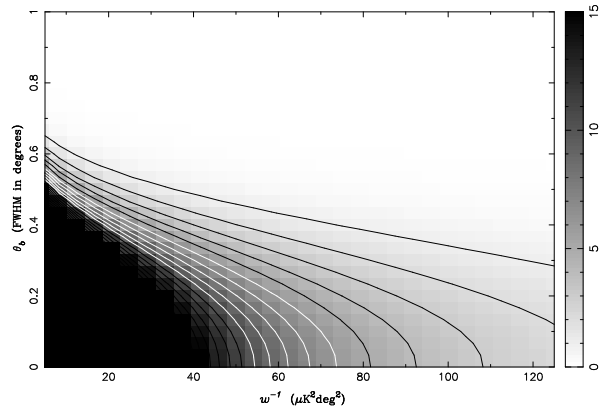,
width=8cm}}
\caption{Contours $\Sigma=1,2,\ldots,10$ for an all-sky experiment.
The noise $w^{-1}$ varies from $(5\mu K)^{2}(\rm{deg})^{2}$
to $(125\mu K)^{2}(\rm{deg})^{2}$. The no-noise axis (not plotted)
is the 76.6 contour which asymptotically absorbs all other contours.}
\label{fig7}
\end{figure}

The first section is the all-sky limit, which may be obtained in our
formalism, not by considering an infinite patch of sky, but by setting
$L^2=4\pi$. This is equivalent to setting $L\approx 202$ degrees.  The
all-sky section, plotted in Fig.~\ref{fig7}, shows first of all that
the main consideration for detecting the secondary peaks is not the
beam size, but the noise level.
In the absence of noise ($w^{-1}=0$) one could in theory deconvolve a
beam of any size without adding any extra uncertainty.  It can be
checked that in the all-sky case the $w^{-1}=0$ axis is the contour
$\Sigma\approx 77$ of the detection function. This is also the maximum
possible detection, imposed by cosmic variance, for an an all-sky
single-dish experiment (with any beam).  As soon as noise is taken
into account, however, this upper limit is greatly reduced, and the
beamsize becomes crucial.  Even for the low (but realistic) level of
noise $w^{-1}=(15\mu K)^{2}(\rm{deg})^{2}$ a modest 1-sigma detection
requires $\theta_b=0.6^{\circ}$, whereas a 3-sigma detection requires
$\theta_b=0.5^{\circ}$.  This openly contradicts the naive argument
favouring any beam with $\theta _b<0.85^{\circ}$. Although such beams do not
start to cut off before $l=1000$, the noise is the real limiting
factor. If the level of noise is pushed up to $w^{-1}=(125\mu
K)^{2}(\rm{deg})^{2}$ an all sky experiment would never provide more than a
2-sigma detection, and even this would require $\theta _b=0.1^{\circ}$.

\begin{figure}
\centerline{\epsfig{
file=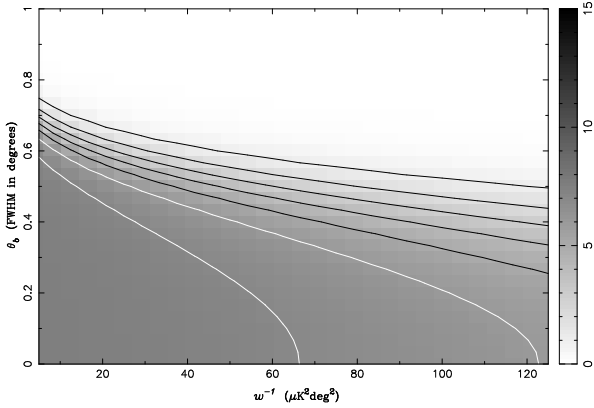,
width=8cm}}
\caption{Same as Fig.~\ref{fig7}, but for a square field with
$L=20$ degrees.}
\label{fig8}
\end{figure}

The all-sky diagram is misleading, however, as for some levels
of noise and beam sizes it may be advantageous to reduce the sky
coverage. In Fig.~\ref{fig8} we plot the detection function for 
$L=20$ degrees. We see that all combinations of noise and beamsize outside
the $\Sigma=6$ contour for an all-sky experiment profit from
reducing the sky coverage. We call this the noise-dominated region,
and its complementary region signal-dominated. 
The effects of noise are always reduced by concentrating all the
observation time on a smaller area, since then we allow the {\it same}
coherent signal to compete with the incoherent noise. It therefore
makes sense that, if one is in the noise-dominated region, the
detection function increases by reducing the area of sky observed.  In
the signal-dominated regions, on the other hand, the detection
function is controlled by the cosmic/sample variance. Different
observers in the Universe looking at different sky patches see
different signals scattered about the theoretical average with a
variance equal to the cosmic/sample variance.  Although we cannot
increase the number of skies observed, we can increase the area of sky
sampled, thereby decreasing the cosmic/sample variance.  For a given
sample size the cosmic/sample variance limit is achieved at the
no-noise axis and is roughly $\Sigma\approx 77 L^2/(4\pi)$.  By
increasing $L$ the cosmic/sample variance decreases and so the
detection function increases in signal-dominated regions.

The contrasting behaviour of the noise- and signal-dominated regions
can be explained by the following semi-quantitative argument. 
From Eqn.~(\ref{varnoise1}) we have roughly
\begin{equation}
\Sigma^2\propto {\Omega^s\over 
{\left(1+{w^{-1}\Omega^s \over  B_i^2C_i}\right)}^2}
\end{equation}
where $C_i$ and $B_i$ are the average power spectrum and beam value
for the bins used in estimating the convexity $\Sigma$.  For a low
signal-to-noise ratio, the noise term in the denominator dominates
and so $\Sigma^2\propto 1/\Omega^s$.  On the other hand, for a high
signal-to-noise ratio, the noise term is much smaller than 1, and so
$\Sigma^2\propto 1/\Omega^s - \beta \Omega^{s2}$, where $\beta\ll 1$.

\begin{figure}
\centerline{\epsfig{
file=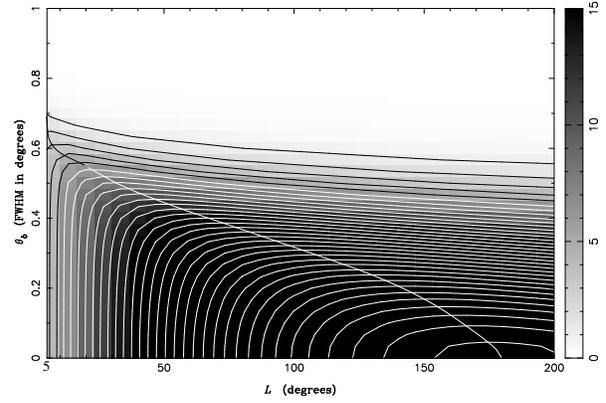,
width=8cm}}
\caption{Low noise ($w^{-1}=(25\mu K)^{2}(\rm{deg})^{2}$) contours
of the $\Sigma$ function. 
We allow the sky coverage $L$ to vary
between 5 degrees and all sky ($L=202$ degrees), 
and consider beamsizes with FWHM between
0 and 1 degrees.}
\label{fig9}
\end{figure}

The question naturally arises as to what is the ideal
scanning strategy for a given experiment, and assuming that ideal
strategy how significant is the detection of the sCDM secondary
Doppler peaks.  The answer to this question depends on the values of
noise level $w^{-1}$, which itself depends on current detector
technology and how effectively foreground emission can be subtracted.
If we assume a low noise value (such as, for definiteness,
$w^{-1}=(25\mu K)^{2}(\rm{deg})^{2}$), then the contours of $\Sigma$ in the
beam-sky coverage plane are as shown in Fig.~\ref{fig9}.
For any beamsize there is a maximum sky coverage beyond which the
detection is not improved.  If anything the level of the detection
decreases, but typically not by much.  The ideal scanning strategy is
then defined by a line $L_i(\theta_b)$ which intersects the contours
of $\Sigma$ at the lowest $L$-value at which a plateau has been
achieved in the detection function.  The significance of the detection
obtained for an ideally scanned experiment depends on the beam
size. For example, if $\theta_b=0.6^{\circ}$, the ideal coverage is a
patch of $L_i(0.6^{\circ})=5$ degrees, which results in a 3-sigma
detection.  If $\theta_b=0.5^{\circ}$, on the other hand, an 8-sigma
detection can be obtained with $L_i=35$ degrees.  The detection
provided by an optimally scanned experiment increases at first very
quickly as the beam is reduced below $\theta_b=0.6^{\circ}$ (from
3-sigma at $\theta=0.6^{\circ}$ to 33-sigma at $\theta_b=
0.2^{\circ}$). By reducing $\theta_b$ from $0.2^{\circ}$ to zero,
however, the detection is only increased by 2-sigma (from 33 to 35).
For this level of noise the maximal detection is 35 sigma and is
achieved with $\theta_b<3'$ and all-sky coverage.  
For low noise levels all-sky coverage
is never harmful, but it
is the beamsize that determines how good a detection can be achieved,
and how much sky coverage is actually required for an optimum level of
detection.

\begin{figure}
\centerline{\epsfig{
file=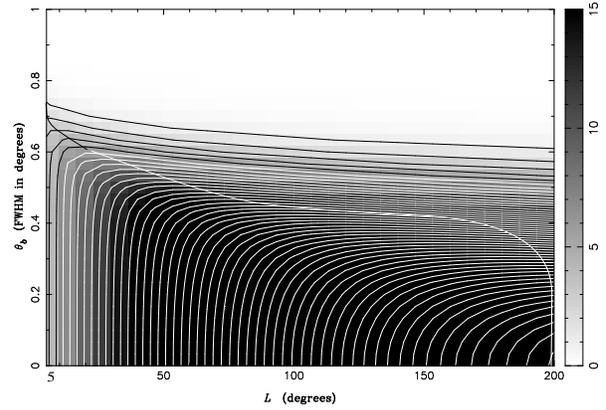,
height=5.5cm,
width=8cm}}
\caption{Very low noise ($w^{-1}=(15\mu K)^{2}(\rm{deg})^{2}$)
contours of the $\Sigma$ function.}
\label{fig10}
\end{figure}

For noise levels of the order $w^{-1}=(25\mu K)^{2}(^o)^{2}$ the
overall picture is always as in Fig.~\ref{fig9}.  In particular, there
is always a top contour (like the $\Sigma=35$ contour in
Fig.~\ref{fig9}) referring to the maximal detection allowed by the
given noise level. The maximum $\Sigma$ is always achieved with
infinite resolution, but one falls short of this maximum by only a
couple of sigmas if $\theta_b\approx 0.1^{\circ}$.  If the noise is
much smaller than this, however, the summit of $\Sigma$ is beyond
$L=202^{\circ}$, as in Fig.~\ref{fig10}. For $w^{-1}=(15\mu
K)^{2}(\rm{deg})^{2}$, for instance, all-sky coverage becomes ideal
for any $\theta_b<0.3^{\circ}$.

\begin{figure}
\centerline{\epsfig{
file=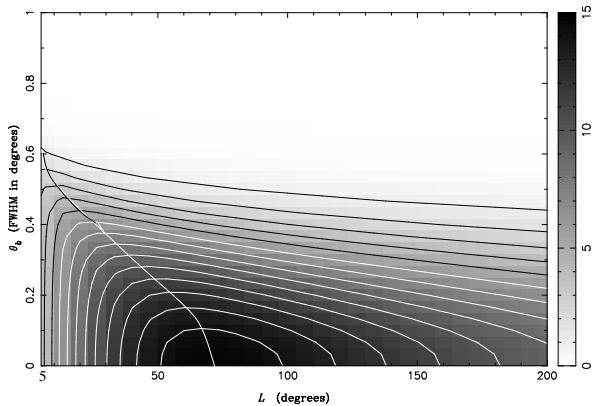,
width=8cm}}
\caption{High noise ($w^{-1}=(60\mu K)^{2}(\rm{deg})^{2}$)
contours of the $\Sigma$ function.}
\label{fig11}
\end{figure}

If, on the other hand, the noise is much larger than $w^{-1}=(25\mu
K)^{2}(\rm{deg})^{2}$ then the $\Sigma$ contours are qualitatively
different from Fig.~\ref{fig9}.  For, say $w^{-1}=(60\mu
K)^{2}(\rm{deg})^{2}$, the contours of $\Sigma$ are shown in
Fig.~\ref{fig11}.  
The beamsize is now a crucial factor. A beamsize of
$\theta_b=0.5^{\circ}$ would provide a 3-sigma detection
(with $L_i=10^{\circ}$), but reducing
the beamsize to about $\theta_b=0.4^{\circ}$ improves the detection to
6-sigma (with $L_i=20^{\circ}$). 
Figure Fig.~\ref{fig11} also shows that, for high noise
levels, forcing all-sky coverage dramatically decreases the detection.

We note finally that in interpreting the figures presented in this
section, we must remember that in order to achieve low values of
$w^{-1}$, the necessity for point source subtraction may severely
limit the maximum possible sky coverage. In turn this can greatly
reduce the significance of possible detections.

\subsection{Interferometers}

For inteferometer observations the detection function
$\Sigma=\Sigma(w^{-1},\theta_w,n_{\rm f})$, so it is a function of the
noise level $w^{-1}$, the size of a single field
(which is characterised by the FWHM of the Gaussian primary beam
$\theta_{\rm w}\approx 2.35\sigma_{\rm w}$), 
and the number of such fields $n_{\rm f}$. Alternatively,
noting that the solid-angle of a single field $\Omega_{\rm s} = \pi
\sigma_{\rm w}^2$
we can instead write the detection function as
$\Sigma=\Sigma(w^{-1},\theta_w,\Omega_{\rm t})$,
where $\Omega_{\rm t}=n_{\rm f}\Omega_{\rm s}$ is the solid angle
of the {\em total} area of sky observed.

For ground-based interferometers it is expected that atmospheric
effects will limit the shortest possible antenna spacing, which in
turn will restrict the largest allowable FWHM of the primary beam
to be $\sim$ 5 degrees. Since this is also close to the limit where
convolution of the power spectrum becomes important for sCDM,
in this section we shall restrict our attention to this case.

\begin{figure}
\centerline{\epsfig{
file=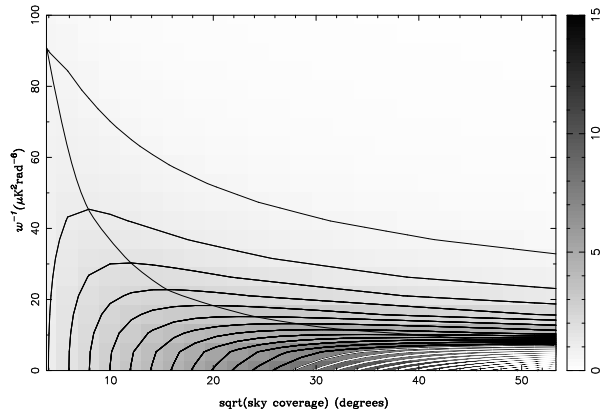,
width=8cm}}
\caption{Contours of the detection function $\Sigma$ 
for an interferometer with a primary beam with $\theta_w=5^{\circ}$, 
for varying noise levels $w^{-1}$
and sky coverage.}
\label{fig12}
\end{figure}

In Fig.~\ref{fig12} we have plotted the contours of 
$\Sigma$ for $\theta_{\rm w}=5^{\circ}$. 
As in the single-dish case, we
we find that for any given noise level there is a sky coverage beyond
which the detection function saturates and then starts to decrease. 
As the noise level decreases, the 
the saturation sky coverage increases along with the
significance of the detection achieved.

In the high-noise r\'egime $w^{-1}>(50\mu K)^2{\rm rad}^{-6}$ the
ideal scanning strategy is to observe only a small sky area (a few
fields at most), which results in only a 1- to 2-sigma detection.
Increasing the sky coverage in the high-noise r\'egime seriously
decreases the level of the detection.  We note that a 1-sigma
dectection requires $w^{-1}<(100\mu K)^2{\rm rad}^{-6}$, and a 2 sigma
detection may be achieved with $w^{-1}=(50\mu K)^2{\rm rad}^{-6}$.
For $w^{-1}<(50\mu K)^2{\rm rad}^{-6}$ we enter the signal dominated
region.  By reducing $w^{-1}$ both the significance of the detection
and the ideal sky coverage (no. of fields) increase very quickly.  We
note that observing just a single field provides only a 2-sigma
detection dictated by cosmic/sample variance. By increasing the number
of observed fields up to a saturation value, however, increases the
level of the detection significantly. In the signal-dominated
r\'egime, increasing the number of fields beyond the ideal value
decreases the detection only very slightly.
For low noise levels $w^{-1}\approx (10\mu K)^2{\rm rad}^{-6}$ 
we see that an 9-sigma detection is possible for an ideal total sky coverage
of $\Omega_{\rm t} \approx$ (45 deg)$^2$. 
We also note that in the low-noise case, detections of several sigma
are possible even for relatively
small sky-coverage. This is an important consideration if source subtraction
is taken into account, since (as mentioned earlier) this severely
limits the total possible sky coverage.

\section{Distinguishing between cosmic strings and inflation}\label{cs}

It is shown in Albrecht et al (1995) and Magueijo et al (1995) that
the power spectrum of cosmic strings does not possess secondary
Doppler peaks. It instead has a single peak at multipoles around
$\ell=400-600$, although the height and precise shape of the peak are
still subject to a considerable theoretical uncertainty.

If future observations of the CMBR show that the main peak in the true CMBR
power spectrum is at lower multipoles, such as $\ell=200-250$ (as
predicted by sCDM), then cosmic strings (in their present form) can be
rejected as a possible theory for structure
formation in the Universe. If, however, the main peak of the true CMBR
power spectrum is shifted towards higher multipoles, then we must rely
on the presence or absense of secondary peaks in order to distinguish
between cosmic strings and CDM/inflation scenarios. 

In this section we consider the case of maximal confusion by comparing
a cosmic strings model and a CDM model for which the main peak in the
power spectrum has the same position and shape (but the latter
exhibits secondary peaks).  For definiteness we have chosen a CDM
theory with a flat primordial spectrum, $\Omega=0.3$, $h=0.6$, and
$\Omega_b h^2=0.02$.  We shall call this theory stCDM, the CDM
competitor of cosmic strings.  We have normalised the power spectra so
that the height and shape of the main peak are similar in both cases.
In principle we could now study the convexities associated with bins
adjusted to the stCDM secondary oscillations, study the same
quantities drawn from a cosmic strings spectrum, and then calculate
the difference between the convexities average values in the two
theories in units of the average variance.  However, this procedure
would define a `distance' between the two theories which depends on
the shape of the strings peak.  We therefore simply study the first
dip detection function of stCDM, and then take this detection function
as a cosmic string rejection function.
 
\begin{figure}
\centerline{\epsfig{
file=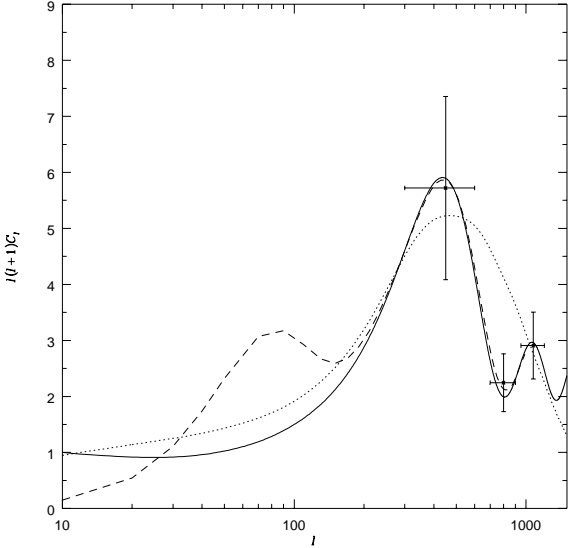,
width=8cm}}
\caption{The angular power spectrum of stCDM (solid line) and one possible
cosmic string scenario (dotted line). The dashed line is the stCDM 
power spectrum $C^s(k)$ as sampled by an interferometer with a primary
of 2 degrees FWHM. The points indicate the average power in each bin
for stCDM. The horizontal errorbars denote the width of the bins,
and the vertical show the sample
variance of the power estimtates for the same interferometer, assuming
no noise.}
\label{fig13}
\end{figure}

In Fig.~\ref{fig13} we show the angular power spectrum of stCDM (solid
line) and a possible power spectrum for cosmic strings (dotted line). 
We also plot
the stCDM  power spectrum $C^s(k)$ (divided by $\alpha$) as sampled by an 
interferometer with a primary beam of 2 degrees FWHM (dashed line). 
We see that, since the secondary
peaks for stCDM are 
now much more separated in $\ell$ space (as compared to sCDM),
the field size required for convolution of the power spectrum 
not to be a problem is much smaller. It can be checked
that interferometers with a primary beam of FWHM $\ga 1.5$ degrees,
and square
windows (multiplied by a cosine bell) with $L \ga 4$ degrees 
are perfectly acceptable.
We therefore assume these values as lower bounds on the field size
for the purposes of cosmic string detection. For fields smaller than this
deconvolution of the power spectrum would be necessary 
causing a deterioration of the detection
function not taken into account in our calculations. However,
we will see that the field must always be larger than
this for any reasonable detection to be achieved, even baring deconvolution
effects.

\begin{figure}
\centerline{\epsfig{
file=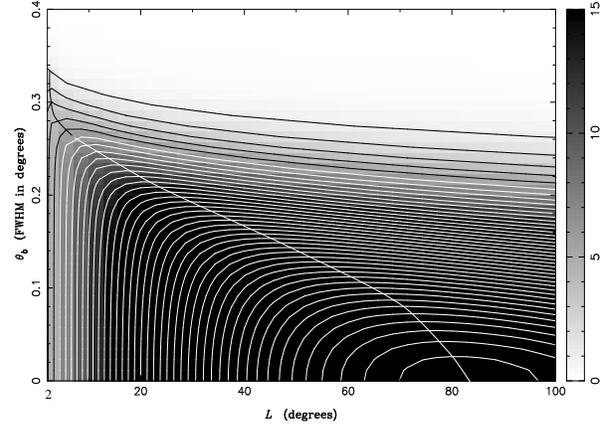,
width=8cm}}
\caption{Low noise $w^{-1}=(25\mu K)^{2}(\rm{deg})^{2}$ contours
of the detection function $\Sigma$ for stCDM.}
\label{fig14}
\end{figure}

\begin{figure}
\centerline{\epsfig{
file=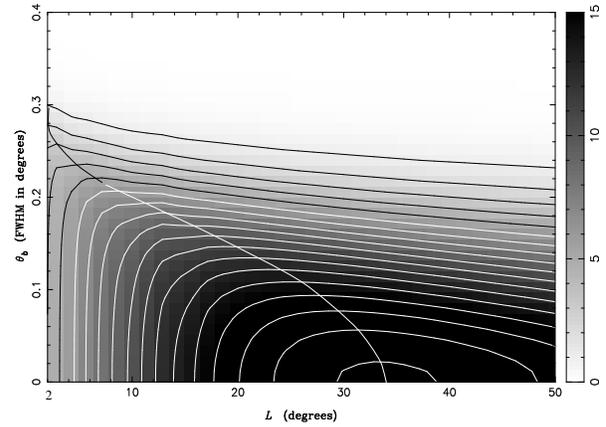,
width=8cm}}
\caption{High noise $w^{-1}=(60\mu K)^{2}(\rm{deg})^{2}$ contours
of the detection function $\Sigma$ for stCDM.}
\label{fig15}
\end{figure}

\begin{figure}
\centerline{\epsfig{
file=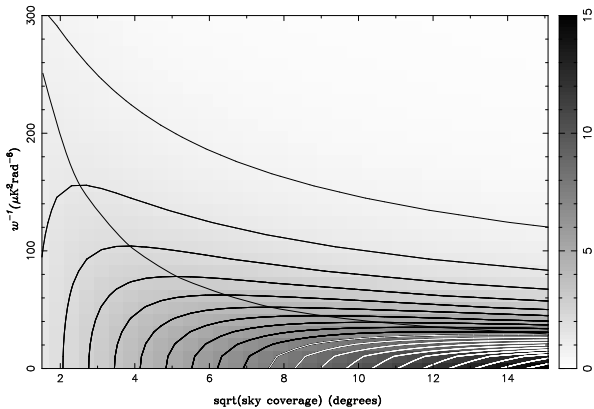,
width=8cm}}
\caption{Contours of the detection function $\Sigma$ for stCDM
for an interferometer with a primary
beam with $\theta_w=2^{\circ}$, for varying noise levels $w^{-1}$
and sky coverage.}
\label{fig16}
\end{figure}

In Fig.~\ref{fig13} we also show
the bins in $\ell$ chosen for the study of the convexity associated with the
first dip in the stCDM power spectrum. The vertical errorbars are the
sample variance errorbars associated with these bins for an 
interferometer with a primary beam of 2 degrees FWHM.
We then repeat the same exercise as in the previous section
concerning the detection function of the first dip of stCDM.
In Figs.~\ref{fig14}, \ref{fig15}, and \ref{fig16}, we redraw the experimental
parameter space covered in Figs.~\ref{fig9},
\ref{fig11}, and \ref{fig12} but as applied to the first dip stCDM 
detection function. The only difference
is that now we allow $L$ to start at 2 degrees. Similar, for
interferometers, we consider a primary beam of 2 degrees FWHM.
Again we make the point that if the
effects of source subtraction are taken into account the total
possible sky coverage may be severely limited, and this should be
borne in mind when considering the figures presented in this section.

Overall we see that in signal dominated regions the detection
is much better for stCDM than for sCDM. This is because features at
higher $\ell$ have a smaller cosmic/sample variance (which
is proportional to $1/\ell$). It can be checked that the cosmic/sample
variance limit, obtained with a single-dish experiment with no noise,
is now $\Sigma\approx 197L^2/(4\pi)$ (as opposed to 
$\Sigma\approx 77L^2/(4\pi)$ for sCDM).
Even in the presence of noise, wherever the signal dominates, the detection
is better for stCDM. However, in noise-dominated regions 
the behaviour of the detection function
for stCDM and CDM
is very different and the changes introduced follow a different logic for 
single-dish and interferometers. 

For single-dish experiments the signal-dominated region is greatly reduced
in stCDM. Much smaller beamsizes $\theta_b$ are now required for any meaningful
detection. As shown in Figs.~\ref{fig14}, and \ref{fig15} one would now
need $\theta_b<0.3^{\circ}$ and $\theta_b<0.25^{\circ}$, for noises
$w^{-1}=(25\mu K)^{2}(\rm{deg})^{2}$ and $w^{-1}=(60\mu K)^{2}(\rm{deg})^{2}$
respectively, in order
to obtain any reasonable  detection. Again one can plot an ideal 
scanning line in the beam/coverage sections defined by a fixed 
noise $w^{-1}$. The ideal sky coverage is much smaller 
for stCDM than for sCDM. In general the countours of 
$\Sigma$ for 
stCDM compared
to sCDM are squashed to lower $\theta_b$, lower $L$, and achieve higher
significance levels, with steeper slopes. Following an ideal scanning line
for any fixed $w^{-1}$ one reaches a maximal 
detection allowed by the given level of noise which is always
better for stCDM than for sCDM. This maximal detection
is normally obtained with a small sky coverage, and infinite
resolution.
Nevertheless, one falls short of this maximum by only a few sigma 
if the resolution is about $\theta_{\rm b} = 4'-6'$. 
From Fig.~\ref{fig14},
for $w^{-1}=(25\mu K)^{2}(\rm{deg})^{2}$, one may now obtain a maximal 43-sigma
detection for an ideal scanning area of $L=65$ degrees. 
If $\theta_b=0.1^{\circ}$ a 36-sigma detection is still obtained.
We also see that a beamsize of $\theta_b < 0.25^{\circ}$ is required
to obtain a 3-sigma detection (with $L=4$ degrees), and  
a 10-sigma detection can be achieved
only with $\theta_b\approx 0.15^{\circ}$ (and $L_i=18$ degrees).
All-sky coverage for an experiment targeting 
stCDM is generally inadvisable, and it would only be optimal 
for the unrealistically low levels of noise 
$w^{-1}<(11\mu K)^{2}(\rm{deg})^{2}$.

For interferometers the signal dominated regions are similar for sCDM
and stCDM (compare Fig.~\ref{fig12} with Fig.~\ref{fig16}). 
The borderline detection regions (where noise dominates), however,
have expanded.  This is because for stCDM we may make the primary beam
smaller without running into convolution problems, which in turn
reduces the instrumental noise.  The high noise region is now
$w^{-1}>(150\mu K)^2{\rm rad}^{-6}$. There one should observe only one
or two $2^{\circ}$ fields, in order to obtain a detection between 1-
and 2-sigma. For $w^{-1}<(150\mu K)^2{\rm rad}^{-6}$ we enter the
signal dominated region. Following the ideal scanning line with
decreasing $w^{-1}$, both the sky coverage and the level of the
detection start increasing, at first first slowly, but then very
quickly. Even for very high noise levels
$w^{-1}=(100\mu K)^2{\rm rad}^{-6}$ one may obtain a 
3-sigma detection with a total sky coverage of about (4 deg)$^2$
(which corresponds to only 7 independent fields)
For relatively low-noise
$w^{-1}=(20\mu K)^2{\rm rad}^{-6}$, on the other hand, 
we may obtain an 8-sigma detection with a sky coverage of around (10
deg)$^2$. Moreover, in the low-noise case, detections
of several sigmas are possible for very small sky coverage.

\section{Conclusions}\label{conc}
We have studied the significance of secondary peak detection
for sCDM and stCDM in a large parameter space of experiments,
including interferometers and single-dish telescopes, and
have adopted a broad minded
attitude towards sky coverage. If point source subtraction is
to be done in parallel with a CMBR experiment (so as to account
for variable point-sources), however, a large sky-coverage may never
be possible (see Section~\ref{point}). 
This detail, often overlooked in satellite proposals,
could then radically undermine a large number of estimates 
assisting experimental design. We will however not dwell on this 
awkward
possibility, but while keeping it in mind, shall consider
unrestricted sky coverage.

The results obtained are reported in Sections~\ref{peakstat} and \ref{cs}
and stress the contradictions of an all-purpose experiment.
If the low-$l$ plateau of the spectrum is a theoretical target
then one needs all-sky coverage, and
satellite single-dish  experiments are to be favoured. 
As shown in Section~\ref{peakstat} even if one wishes to study 
Doppler peak features for sCDM 
all-sky coverage might still be preferable.
Depending on the noise levels, a large sky coverage might be desirable,
even for a resolution of about $\theta_b=0.4-0.5^{\circ}$.

Our work shows how such a design relies heavily on the assumption that
the signal is in the vicinity of sCDM. If instead one is to test
the high-$l$ opposition between low $\Omega$ CDM and cosmic strings, 
then we have seen that single-dish experiments are
required to have rather high resolutions. 
Interferometers appear to be less constrained,
providing 2-3 sigma detections under very unassuming conditions,
with rapid improvements following further 
improvement
in experimental conditions.
Furthermore, in this context, all-sky scanning is not only
unnecessary, but in fact undesirable. 
The best scanning
is normally achieved with deep small patches. These two features
contradict sharply the ideal experimental design motivated by the standard
theoretical gospel. 

Overall the parameter space of successful stCDM detections seems
to increase for interferometers and shrink for single-dish
when compared with sCDM. 
We believe that a variety of contrasting experimental
techniques may equally well find their niche as regards
important theoretical
implications.

\section*{Acknowledgments}

The authors would like to thank A.~Albrecht, P.~Ferreira, M.~Jones,
and B.~Wandelt for useful discussions. We also thank M.~White for
supplying the stCDM $C^l$ spectrum. We acknowledge  
Trinity Hall (MPH) and St.~John's College (JM),
Cambridge, for support in the form of Research Fellowships.

\bsp  
\label{lastpage}

\end{document}